\documentclass{SciPost}

\hypersetup{
    colorlinks,
    linkcolor={red!50!black},
    citecolor={blue!50!black},
    urlcolor={blue!80!black}
}
\newcommand{\magenta}[1]{\color{magenta} #1 \color{black}}

\usepackage[bitstream-charter]{mathdesign}
\usepackage{cancel}
\usepackage{multicol}
\usepackage{multirow}
\usepackage{amsmath}
\usepackage{slashed}
\usepackage{dsfont}
\usepackage[normalem]{ulem}
\usepackage{hyperref}

\DeclareSymbolFont{usualmathcal}{OMS}{cmsy}{m}{n}
\DeclareSymbolFontAlphabet{\mathcal}{usualmathcal}

\fancypagestyle{SPstyle}{
\fancyhf{}
\lhead{\colorbox{scipostblue}{\bf \color{white}Prepared for submission to SciPost Physics Community Reports }}
\rhead{{\bf \color{scipostdeepblue}}}

\fancyfoot[C]{\textbf{\thepage}}
}

\newcommand{\be}{\begin{equation}}
\newcommand{\ee}{\end{equation}}
\newcommand{\bei}{\begin{itemize}}
\newcommand{\eei}{\end{itemize}}

\newcommand{\cL}{{\mathcal L}}

\newcommand{\nnl}{\nonumber \\}  

\def\hc{{\rm h.c.}}

\begin{document} 
\vspace*{-1cm}
\begin{flushleft} 
\magenta{LHCHWG-2025-004}\\
\magenta{IFT-UAM/CSIC-25-147}
\end{flushleft}

\pagestyle{SPstyle}

\begin{center}{\Large \textbf{\color{scipostdeepblue}{
Parametrisation and dictionary for CP violating Higgs boson interactions
}}}\end{center}

\begin{center}\textbf{
Daniele Barducci\textsuperscript{1},
Matthew Forslund\textsuperscript{2,3,4},
Marta Fuentes Zamoro\textsuperscript{5},\\
Pier Paolo Giardino \textsuperscript{5},
Andrei V. Gritsan\textsuperscript{6}
and Giacomo Ortona\textsuperscript{7}  
}\end{center}

\begin{center}
{\bf 1} Dipartimento di Fisica ``Enrico Fermi'', Universit\`a di Pisa and INFN Sezione di Pisa, Largo Bruno Pontecorvo 3, I-56127,
Pisa, Italy
\\
{\bf 2} C. N. Yang Institute for Theoretical Physics, Stony Brook University, Stony Brook, NY, 11794, USA
\\
{\bf 3} High Energy Theory Group, Physics Department, Brookhaven National Laboratory, Upton, NY 11973, USA
\\
{\bf 4} Princeton Center for Theoretical Science, Princeton University, Princeton, NJ, 08544 USA
\\
{\bf 5} Departamento de F\'{i}sica Te\'{o}rica and Instituto de F\'{i}sica Te\'{o}rica UAM/CSIC, Universidad Aut\'{o}noma de Madrid, Cantoblanco, 28049, Madrid, Spain
\\
{\bf 6} Department of Physics and Astronomy, Johns Hopkins University, Baltimore, MD 21218, USA
\\
{\bf 7} Istituto Nazionale di Fisica Nucleare, Sezione di Torino, via P. Giuria 1, I–10125 Torino,
Italy
\\[\baselineskip]
 \href{mailto:daniele.barducci@pi.infn.it}{\small daniele.barducci@pi.infn.it}\,,\quad
\href{mailto:mforslund@princeton.edu}{\small mforslund@princeton.edu}\,,\quad
 \href{mailto:marta.zamoro@uam.es}{\small marta.zamoro@uam.es}\,,\quad
 \href{mailto:gritsan@jhu.edu}{\small gritsan@jhu.edu}\,,\quad
  \href{mailto:giacomo.ortona@to.infn.it}{\small giacomo.ortona@to.infn.it}\,,\quad
 \href{mailto:pier.giardino@uam.es}{\small pier.giardino@uam.es}
\end{center}

\section*{\color{scipostdeepblue}{Abstract}}
\textbf{\boldmath{Searches for charge-parity (CP) violating interactions of the Standard Model (SM) Higgs boson are a key priority of the LHC physics program. Experimental results from ATLAS and CMS are often reinterpreted within a variety of theoretical parametrisations, the most commonly used being the Higgs basis, $\kappa$'s and angles, CP fractions and effective field theories (EFT) such as the SMEFT and the Higgs EFT. However, differing conventions and assumptions across the literature make the translation between these parametrisations nontrivial and prone to inconsistencies.
In this paper, we provide a unified framework and construct explicit dictionaries connecting these different approaches. This facilitates a transparent comparison between theoretical studies and experimental analyses, enabling more robust interpretations of CP violating effects in Higgs boson interactions.
}}

\vspace{\baselineskip}




\vspace{10pt}
\noindent\rule{\textwidth}{1pt}
\tableofcontents
\noindent\rule{\textwidth}{1pt}
\vspace{10pt}


\section{Introduction \label{sec:intro}}

Additional sources of charge-parity (CP) symmetry violation, beyond the complex phase of the CKM matrix in the Standard Model (SM), are a necessary condition for explaining the observed baryon asymmetry of the Universe (BAU) \cite{Sakharov:1967dj}. Among the various possibilities, CP violating (CPV) interactions involving the Standard Model Higgs boson are of particular interest, as they offer a potential mechanism for BAU generation through electroweak (EW) baryogenesis~\cite{Kuzmin:1985mm,Shaposhnikov:1986jp,Shaposhnikov:1987tw,Morrissey:2012db}. In brief, this mechanism relies on making the electroweak phase transition strongly first-order, a condition typically achieved through the presence of new physics (NP) in the scalar sector. During the phase transition, expanding bubbles of the true vacuum nucleate within the false vacuum background. At the bubble walls, CP violating Higgs boson interactions generate a chiral asymmetry in front of the wall, which is then partially converted into a baryon asymmetry by EW sphaleron processes active in the unbroken phase. As the bubbles expand, the resulting baryon excess becomes trapped inside, where sphaleron processes are suppressed, thus protecting the generated asymmetry from washout.

For this reason the search of CP violating effects involving the SM Higgs boson is among the priorities of the LHC and HL-LHC program, also motivated by a more agnostic view of scrutinising at best the properties of the last elementary particle so far discovered\,\cite{ATLAS:2012yve,CMS:2012qbp}. 
In particular, the ongoing analyses at the LHC and the $3\,$ab$^{-1}$ of data expected at the HL-LHC will play a decisive role in directly constraining CP violating Higgs interactions. These measurements will provide the most sensitive and comprehensive tests of the Higgs sector achievable before the advent of future collider facilities, and will serve as the reference benchmark for subsequent studies.
The ATLAS and CMS experiments have performed several analyses for the direct search of CP violating Higgs boson couplings, testing interactions with EW bosons\,\cite{ATLAS:2023mqy,ATLAS:2022tan,CMS:2024bua,ATLAS:2025hki,CMS:2021nnc,CMS:2012vby,CMS:2013fjq,CMS:2014nkk,CMS:2015chx,CMS:2016tad,CMS:2017len,CMS:2019ekd,CMS:2019jdw,ATLAS:2013xga,ATLAS:2015zhl,ATLAS:2016ifi,ATLAS:2017qey,ATLAS:2017azn,ATLAS:2018hxb,ATLAS:2020evk,CERN-EP-2025-127,CMS:2022uox,ATLAS:2024wfv,ATLAS:2025bts}, gluons\,\cite{ATLAS:2021pkb,CMS:2021nnc,CMS:2022uox,CMS:2024bua}, top quarks\,\cite{CMS:2020cga,CMS:2022dbt,ATLAS:2023cbt,CMS:2024fdo,CMS:2021nnc,ATLAS:2020ior} and $\tau$ leptons\,\cite{CMS:2021sdq,ATLAS:2022akr} through the use of CP-sensitive observables, {\it i.e.} observables sensitive to 
the interference of CP-even and CP-odd contributions.
Also CP-insensitive observables, {\it i.e.} sensitive to CP-even quadratic terms  $\propto |\mathcal{M}_{\text{CP}}|^2$ but not to CP-odd interference terms, can still constrain CP-odd interactions and, in some cases, outperform\footnote{This is with the caveat that bounds from CP-insensitive observables often violate EFT powercounting and cannot be included in global EFT fits -- see Section~\ref{subsec:kappas} for a discussion.} indirect limits arising from the search of electric dipole moments.
This is the case for CP violating Higgs boson interactions with muons\,\cite{Fuchs:2019ore,Bahl:2022yrs}, for example.

While experimental searches measure amplitudes, theoretical models are built from Lagrangians and it is thus necessary to interpret the results of the former in terms of the parameters of the latter. 
Different parametrizations for CP-odd interactions are present in the literature.
For what concerns CP violating Higgs boson couplings, the most common are the ones written in the Higgs basis, through the use of 
coupling modifiers ($\kappa$'s), angles, CP fractions, and in terms of the SM effective field theory (SMEFT) with various basis choices  or the Higgs EFT (HEFT) frameworks. 
We will define and review all of these in the following sections. 
However, for each of these parametrizations different conventions are often defined which can easily lead to mistakes when translating results between them. 
Given the importance and interest of both the experimental and theory community on the topic, this work aims at establishing a parametrization accompanied with a dictionary for CP violating Higgs boson interactions to facilitate the comparison between different studies. 

The paper is organised as follows. In Sec.\,\ref{sec:exp} we briefly review the current status and future prospects for experimental searches for CP violation. Then
in Secs.\,\ref{subsec:anomalous_couplings}-\ref{subsec:HEFT} we set our notation and review the parametrizations for CP violating Higgs boson coupling in the various formalism, while in Sec.\,\ref{subsec:dictionaries} we provide the dictionary between the various parametrisations. We conclude in Sec.\,\ref{sec:conclusions}.

\section{Experimental status \& prospects}\label{sec:exp}

Characterizing the CP properties of the Higgs boson has been a priority for the community since its discovery~\cite{Gao:2010qx,Bolognesi:2012mm,Englert:2012xt,CMS:2012vby,ATLAS:2013xga,Artoisenet:2013puc}, on both the experimental and theoretical fronts. 
As a consequence, a significant effort has been made into developing tools and designing experimental searches that could better target CP violating observables.
Here we will briefly summarize some of the recent literature relevant for CP violation in the Higgs sector.

At the LHC, both CMS and ATLAS have performed analyses in a number of different channels. 
Although the majority of studies use CP fractions and the $\kappa$-framework, recently ATLAS has published results constraining SMEFT operators~\cite{ATLAS:2020rej}. From the theoretical side, several studies have then used the information from experiments to place bounds on EFT coefficients, see {\it{e.g.}}\,\cite{Fuchs:2020uoc,Barman:2021yfh,Bahl:2021dnc,Martini:2021uey,Bahl:2023qwk,Miralles:2024huv} for some recent studies.

Searches for Higgs boson CP violation will also be a priority for any future collider that will be built for the post HL-LHC era.  A comprehensive overview of the prospects for characterizing anomalous Higgs boson couplings to fermions and bosons at future colliders can be found in Ref.~\cite{Gritsan:2022php}. Below, we briefly summarize the main results, referring the reader to the original publication and its references for a more detailed discussion.\footnote{
For recent studies in the context of future accelerators see also Refs. \cite{Mlynarikova:2023bvx,ATLAS:2022hsp} for HL-LHC, Ref.\cite{FCC:2025fub} for FCC-hh, Refs.\cite{Vukasinovic:2024ftu,Li:2024mxw} for the International Linear Collider (ILC), Ref.\cite{Selvaggi:2025kmd} for FCC-ee,  Ref.\cite{Sha:2022bkt} for CEPC, and Refs.\cite{Cassidy:2023lwd,Ruhdorfer:2024dgz,Gurkanli:2025mio} for the muon collider.}

At hadron colliders such as the HL-LHC and FCC-hh, $Hgg$ coupling can be probed through the scattering topology of Higgs boson production in association with two hadronic jets. Indeed, while direct access to gluon polarization in the decay $H \to gg$ is essentially impossible, the CPV characteristics of the $Hgg$ coupling can still be constrained using production kinematics. Furthermore, by combining multiple production and decay channels, it is possible to gain sensitivity to the CPV components of various couplings, including $H\gamma\gamma$, $HZ\gamma$, $HZZ$, and $HWW$.
Among the fermionic couplings, only $Htt$ and $H\tau\tau$ can be accessed at hadron colliders with reasonable sensitivity. The measurement of Higgs boson couplings to the first two fermion families is essentially out of reach in this context, due to the absence of polarization-sensitive observables. The $Hbb$ coupling, while kinematically accessible, cannot be probed for CP structure either, as polarization measurements of the $b$ quarks would lead to a drastic loss in statistics. 

In contrast, $e^+ e^-$ colliders offer significantly enhanced capabilities for studying the CP structure of Higgs boson couplings. Almost all vertices, including those involving gauge bosons and fermions, can be investigated in detail, with the notable exceptions of $Hgg$, which remains inaccessible due to the inability to probe gluon polarization in $H \to gg$, as well as $Htt$, which requires a high enough collider energy to reach the $t\bar{t}H$ kinematic threshold. Overall sensitivities at $e^+ e^-$ colliders are expected to be higher than those at the HL-LHC, with improved control over systematics and backgrounds.

Lepton–hadron colliders  provide another promising avenue, particularly due to their reduced pile-up environment relative to hadron colliders. This cleaner setting allows for more precise measurements and enhances sensitivity to certain couplings.
Moving beyond conventional collider designs, photon–photon colliders present a unique opportunity. In this setup, the Higgs boson can be produced directly via $\gamma\gamma \to H$. This configuration allows for a highly distinctive probe of the $H\gamma\gamma$ coupling: by exploiting the polarization of the photon beams, one can disentangle the CP-even and CP-odd components of the interaction.

Finally, muon colliders are proposed as a powerful option for the post-LHC era. In principle, runs at the Higgs boson resonance (i.e., $\sqrt{s} \simeq m_h$) would allow for a clean study of the CP properties of the $H\mu\mu$ vertex through polarized $\mu^+\mu^-$ annihilation. Alternatively, at higher center-of-mass energies, a muon collider would be well suited for probing $HVV$ and $Ht\bar{t}$ couplings with high precision.

Apart from the obvious need to study the CP violating properties of the Higgs boson at colliders, some insight can be gained by studying low energy observables, such as the electron dipole moment (EDM), where the Higgs  boson enters the process at loop level and provides indirect sensitivity to Higgs boson CP violation. 
A significant increase in precision for the electron EDM is expected within the next years\,\cite{Alarcon:2022ero} and higher order corrections in theoretical calculations will become necessary\,\cite{Brod:2023wsh}. Indirect limits can  complement direct limits from collider searches, see, {\it e.g.},\,\cite{Panico:2018hal,Bahl:2022yrs,Brod:2022bww,Kley:2021yhn} for recent works on the subject, where however one has to recall that  indirect limits obtained from electron EDMs measurement strongly depend on the assumptions about the Yukawa coupling to the first generation and the absence of other CP violating NP effects.

\section{Parametrizations for CPV in Higgs boson interactions \label{sec:parametrisations}}
In the study of the CPV properties of the Higgs boson, different parametrizations have been used. 
Experimentally, CP fractions, $\kappa's$, and angles are typically used~\cite{CMS:2021sdq,ATLAS:2020rej,CMS:2022uox}, which usually have an immediate correspondence in the Higgs basis.
Phenomenologically, studies often use the  SMEFT~\cite{Cirigliano:2016njn,Cirigliano:2016nyn,Brehmer:2017lrt,Bernlochner:2018opw,Englert:2019xhk,ATLAS:2020rej,DasBakshi:2020ejz,Biekotter:2020flu,Fuchs:2020uoc,Bakshi:2021ofj,Degrande:2021zpv,Asteriadis:2024xuk,Asteriadis:2024xts}, most commonly in the Warsaw basis, and some studies have been performed using the more general HEFT~\cite{Gavela:2014vra,Hierro:2015nna,Bhardwaj:2023ufl,Bhardwaj:2024lyr}.
More details of the theoretical framework behind each of them are given below.

\subsection{Higgs basis}\label{subsec:anomalous_couplings}

Our baseline will be the Higgs basis~\cite{Falkowski:2001958,LHCHiggsCrossSectionWorkingGroup:2016ypw,Azatov:2022kbs}, in which interactions are parameterized using the mass eigenstates of the SM.\footnote{Note that there are some small differences in the conventions of~\cite{LHCHiggsCrossSectionWorkingGroup:2016ypw,Azatov:2022kbs}. We will use the notation in~\cite{LHCHiggsCrossSectionWorkingGroup:2016ypw}.}
Here we are interested only in terms with CPV, and so we will ignore interactions that cannot be CP violating. 
Moreover, we will consider only operators with up to four fields.
The full correspondence, including also operators that are not CP violating, may be found at the aforementioned references.
Starting with the gauge field self interactions, we have
\begin{align}\label{eq:HiggsBasis:tgc}
\begin{split}
 \Delta \mathcal{L}_\text{tgc}^\text{CPV} =& \  ie \widetilde{\kappa}_\gamma \widetilde{A}_{\mu\nu} W_\mu^+ W_\nu^- +ig_Lc_\theta \widetilde{\kappa}_z \widetilde{Z}_{\mu\nu} W_\mu^+ W_\nu^- + i \frac{e}{m_W^2}\widetilde{\lambda}_\gamma W_{\mu\nu}^+ W_{\nu\rho}^- \widetilde{A}_{\rho\mu} \\&+ i \frac{g_L c_\theta }{m_W^2} \widetilde{\lambda}_z W_{\mu\nu}^+ W_{\nu\rho}^- \widetilde{Z}_{\rho\mu} + \frac{g_s^3}{v^2}\widetilde{c}_{3g} f^{abc} \widetilde{G}_{\mu\nu}^a G_{\nu\rho}^b G_{\rho\mu}^c \, ,
\end{split}
\end{align}
where $e$, $g_L$, and $g_s$ stand for the QED, $SU(2)_L$ and QCD gauge couplings, respectively, and $c_\theta = \cos\theta_W$, with $\theta_W$ the Weinberg angle. 
The field strength tensor for the $W^\pm$, $Z$, and $A$ are defined by $X_{\mu\nu} = \partial_\mu X_\nu - \partial_\nu X_\mu$, while that for the gluons is $G_{\mu\nu}^a = \partial_\mu G_\nu^a-\partial_\nu G_\mu^a +g_s f^{abc}G_\mu^bG_\nu^c$. The dual gauge fields are defined as usual, $\widetilde{V}_{\mu\nu} = \frac{1}{2}\epsilon_{\mu\nu\rho\sigma} V^{\rho\sigma}$.
CP violation in the triple gauge interactions is controlled by the five coefficients
\begin{equation}
    \widetilde{\kappa}_\gamma, \ \widetilde{\kappa}_z, \ \widetilde{\lambda}_\gamma, \ \widetilde{\lambda}_z, \ \widetilde{c}_{3g} \ .
\end{equation}
There are also anomalous quartic gauge interactions, which read
\begin{align}\label{eq:HiggsBasisQGCs}
\begin{split}
    \Delta \mathcal{L}_{\text{qgc}}^{\text{CPV}} =& -\frac{g_L^2}{2m_W^2}\widetilde{\lambda}_{z}\left(W^+_{\mu\nu} \widetilde{W}^{-}_{\nu\rho} - W^-_{\mu\nu} \widetilde{W}^+_{\nu\rho}\right)\left(W^+_\mu W^-_\rho - W^-_\mu W^+_\rho\right)    \\
    & - \frac{g_L^2 c^2_\theta }{m_W^2} \widetilde{\lambda}_{z} \left[W^+_\mu \left(Z_{\mu\nu} \widetilde{W}^-_{\nu\rho} - {W}^-_{\mu\nu} \widetilde{Z}_{\nu\rho}\right)Z_\rho - W^-_\mu \left(Z_{\mu\nu} \widetilde{W}^+_{\nu\rho} - {W}^+_{\mu\nu} \widetilde{Z}_{\nu\rho}\right)Z_\rho\right] \\
    &-\frac{e^2}{m_W^2} \widetilde{\lambda}_{z} \left[W^+_\mu \left(A_{\mu\nu} \widetilde{W}^-_{\nu\rho} - {W}^-_{\mu\nu} \widetilde{A}_{\nu\rho}\right)A_\rho - W^-_\mu \left(A_{\mu\nu} \widetilde{W}^+_{\nu\rho} - {W}^+_{\mu\nu} \widetilde{A}_{\nu\rho}\right)A_\rho\right]\\
    & - \frac{e g_L c_\theta }{m_W^2} \widetilde{\lambda}_{z} \left[W^+_\mu \left(A_{\mu\nu} \widetilde{W}^-_{\nu\rho} - {W}^-_{\mu\nu} \widetilde{A}_{\nu\rho}\right)Z_\rho - W^-_\mu \left(A_{\mu\nu} \widetilde{W}^+_{\nu\rho} - {W}^+_{\mu\nu} \widetilde{A}_{\nu\rho}\right)Z_\rho\right]\\
    & - \frac{e g_L c_\theta }{m_W^2} \widetilde{\lambda}_{z} \left[W^+_\mu \left(Z_{\mu\nu} \widetilde{W}^-_{\nu\rho} - {W}^-_{\mu\nu} \widetilde{Z}_{\nu\rho}\right)A_\rho - W^-_\mu \left(Z_{\mu\nu} \widetilde{W}^+_{\nu\rho} - {W}^+_{\mu\nu} \widetilde{Z}_{\nu\rho}\right)A_\rho\right]
\end{split}
\end{align}
where gauge invariance restricts their coefficients to be related to $\widetilde{\lambda}_z$.
These interactions must be written out explicitly due to our definition of the field strength tensor for the $SU(2)_L$ gauge fields as $X_{\mu\nu} = \partial_\mu X_\nu - \partial_\nu X_\mu$.

Moving on to the interactions of the SM gauge bosons with the Higgs boson, we have
\begin{eqnarray}\label{eq:HiggsBasis:hvv}
  \Delta \cL_\text{hvv}^\text{CPV}  &= & {h \over v} \left [\widetilde c_{ww}  {g_L^2 \over  2} W_{\mu \nu}^+   \widetilde W_{\mu\nu}^-  
   +  \widetilde c_{gg} {g_s^2 \over 4} G_{\mu \nu}^a \widetilde G_{\mu \nu}^a  
  + \widetilde c_{\gamma \gamma} {e^2 \over 4} A_{\mu \nu} \widetilde A_{\mu \nu} \right . \nnl & & \left . \qquad
  + \widetilde c_{z \gamma} {e g_L \over  2 c_\theta} Z_{\mu \nu} \widetilde A_{\mu\nu}
  + \widetilde c_{zz}  {g_L^2  \over  4 c_\theta^2} Z_{\mu \nu} \widetilde Z_{\mu\nu}
  \right ],
\end{eqnarray}
where $W^\pm_{\mu\nu}$, $A_{\mu\nu}$ 
and $Z_{\mu\nu}$ are the field strengths tensors of the physical EW gauge bosons and $\tilde{X}_{\mu\nu}\equiv \frac{1}{2} \epsilon_{\mu\nu\rho\sigma} {X}^{\rho\sigma}$, with $\epsilon^{0123}=1$, their associated dual fields. The Lagrangian in Eq.\,\eqref{eq:HiggsBasis:hvv} is
controlled by the five parameters:
\begin{eqnarray}
\label{eq:ind} 
&   \widetilde c_{gg}, \ \widetilde c_{zz},  \   \widetilde c_{\gamma \gamma}, \  \widetilde c_{z \gamma}, \   \widetilde c_{ww} \, .
\end{eqnarray}  
Next, we have the Higgs boson couplings to a pair of fermions, described by
\begin{equation}
\label{eq:HiggsBasis:hff}
\Delta \cL_{\rm hff}     =  - {h \over v} \sum_{f \in u,d,e} \sum_{ij}\sqrt{m_{f_i} m_{f_j}}[\delta y_f]_{ij} \, \bar{f}_{R,i} f_{L,j}  + \hc \, ,
\end{equation}
involving the additional parameters
\begin{equation}
\delta y_u, \   \delta y_d,  \ \delta y_e ,
\end{equation}
where each $\delta y_f$ is a $3\times 3$ complex matrix with CPV phases.

The operators with two Higgs fields and a pair of vector bosons or fermions are very similar to those above and read
\begin{align}\label{eq:HiggsBasis:h2}
\begin{split}
    \Delta \mathcal{L}_{{\rm h}^2}^\text{CPV} = &\ {h^2 \over 2v^2} \bigg[\widetilde c^{(2)}_{ww}  {g_L^2 \over  2} W_{\mu \nu}^+   \widetilde W_{\mu\nu}^-  
   +  \widetilde c^{(2)}_{gg} {g_s^2 \over 4} G_{\mu \nu}^a \widetilde G_{\mu \nu}^a  
  + \widetilde c^{(2)}_{\gamma \gamma} {e^2 \over 4} A_{\mu \nu} \widetilde A_{\mu \nu} \\ & \qquad
  + \widetilde c^{(2)}_{z \gamma} {e g_L \over  2 c_\theta} Z_{\mu \nu} \widetilde A_{\mu\nu}
  + \widetilde c^{(2)}_{zz}  {g_L^2  \over  4 c_\theta^2} Z_{\mu \nu} \widetilde Z_{\mu\nu}  \\ & \qquad 
  - \sum_{f \in u,d,e} \sum_{ij}\sqrt{m_{f_i} m_{f_j}}[\delta y^{(2)}_f]_{ij} \, \bar{f}_{R,i} f_{L,j}  + \hc \bigg].
\end{split}
\end{align}
The gauge-fermion vertices may also be modified as
\begin{eqnarray}
\label{eq:HiggsBasis:vff}
\Delta \cL_{{\rm vff}}   \ &=&  \frac{g_L}{\sqrt{2}}  W_\mu^+   \left (\delta g_L^{Wq} \bar{u}_L \gamma^\mu d_L +\delta g_R^{Wq}\bar{u}_R \gamma^\mu d_R + \delta g_L^{W\ell} \bar{\nu}_L \gamma^\mu e_L  \right )
 + \hc
\nnl 
&&+ \sqrt{g_L^2 + g_Y^2} Z_\mu 
\left [ \sum_{f = u,d,e,\nu}    \delta g^{Zf}_L  \bar f_L \bar \gamma^\mu f_L  +  
\sum_{f = u,d,e}   \delta g^{Zf}_R  \bar{f}_R \gamma^\mu \bar f_R  \right ] ,
\end{eqnarray}
where $\delta g_R^{Wq}$ is a $3 \times 3$ complex matrix while the others are $3\times 3$ hermitian matrices.

Additional $HVff$ contact interactions between the Higgs boson, the $Z/W$ boson and the various fermionic currents of the SM are also permitted at $d=5$ and are described by
\begin{eqnarray}
\label{eq:HiggsBasis:hvff}
\Delta \cL_{{\rm hvff}}   \ &=&  \sqrt 2  g_L {h \over v}   W_\mu^+   \left (\delta g_L^{hWq} \bar{u}_L \gamma^\mu d_L +\delta g_R^{hWq}\bar{u}_R \gamma^\mu d_R + \delta g_L^{hW\ell} \bar{\nu}_L \gamma^\mu e_L  \right )
 + \hc
\nnl 
&&+ 2 {h \over v}  \sqrt{g_L^2 + g_Y^2} Z_\mu 
\left [ \sum_{f = u,d,e,\nu}    \delta g^{hZf}_L  \bar f_L \bar \gamma^\mu f_L  +  
\sum_{f = u,d,e}   \delta g^{hZf}_R  \bar{f}_R \gamma^\mu \bar f_R  \right ] ,
\end{eqnarray}
weighted by 8 potentially CPV $3\times 3$ Hermitian matrices and one $3 \times 3$ complex matrix ($\delta g_R^{hWq}$):
\begin{equation}
\delta g^{hW \ell }_L, \delta g^{hWq}_L, \delta g^{hWq}_R, \delta g^{hZu}_L , \delta g^{hZd}_L , \delta g^{hZe}_L , \delta g^{hZu}_R , \delta g^{hZd}_R , \delta g^{hZe}_R \ .
\end{equation}
These can play an important role in EW Higgs boson production, and they also contribute to $H\to 4f$ partial widths. 
This can be further extended to two Higgs boson insertions with corresponding couplings $g_{L/R}^{h^2 V f}$, though these would have 5 fields and so we will not write them explicitly.\footnote{To fix normalization, if these are included they should be written as $\Delta \mathcal{L}_{{\rm{h^2vff}}}\sim(h^2/v^2)\Delta \mathcal{L}_{{\rm vff}}$ for the simple matching to the SMEFT in section~\ref{subsec:dictionaries}.}

Moving on to dipole operators, we have 
\begin{eqnarray}
\label{eq:HiggsBasis:dipoles}
\Delta\mathcal{L}_\text{dipoles} & = &-\frac{1}{4v}\bigg[g_s \sum_{f\in u,d} \frac{\sqrt{m_{f_i} m_{f_j}}}{v} \left([d_{Gf}]_{ij}+\frac{h}{v}[d_{hGf}]_{ij}\right)\left( \bar{f}_{L,i} \sigma^{\mu\nu}T^af_{R,j} G^a_{\mu\nu} \right)
\nnl 
&&+ e\sum_{f \in u,d,e}\frac{\sqrt{m_{f_i} m_{f_j}}}{v}\left([d_{Af}]_{ij}+\frac{h}{v}[d_{hAf}]_{ij}\right)\left( \bar{f}_{L,i} \sigma^{\mu\nu} f_{R,j} A_{\mu\nu} \right)
\nnl
&&+ \sqrt{g_L^2+g_Y^2} \sum_{f \in u,d,e}\frac{\sqrt{m_{f_i} m_{f_j}}}{v} \left([d_{Zf}]_{ij}+\frac{h}{v}[d_{hZf}]_{ij}\right)\left( \bar{f}_{L,i} \sigma^{\mu\nu}f_{R,j} Z_{\mu\nu}  \right)
\nnl 
&&+ \sqrt{2}g_L \frac{\sqrt{m_{u_i} m_{u_j}}}{v} \left([d_{Wu}]_{ij}+\frac{h}{v}[d_{hWu}]_{ij}\right)\left( \bar{d}_{L,i} \sigma^{\mu\nu} u_{R,j} W^{-}_{\mu\nu}\right)
\nnl
&&+ \sqrt{2}g_L \frac{\sqrt{m_{d_i} m_{d_j}}}{v}\left([d_{Wd}]_{ij}+\frac{h}{v}[d_{hWd}]_{ij}\right)\left( \bar{u}_{L,i} \sigma^{\mu\nu} d_{R,j} W^{+}_{\mu\nu}\right)
\nnl 
&&+ \sqrt{2}g_L \frac{\sqrt{m_{e_i} m_{e_j}}}{v} \left([d_{We}]_{ij}+\frac{h}{v}[d_{hWe}]_{ij}\right)\left( \bar{\nu}_{L,i} \sigma^{\mu\nu} e_{R,j} W^{+}_{\mu\nu}\right)
\bigg] + \hc \ ,
\end{eqnarray}
where the coefficients are complex $3\times 3$ matrices\footnote{Note that these coefficients also generate $\sim \bar{f} f X_\mu X'_\nu$ interactions from our definition of $X_{\mu\nu}$ in a similar manner to the anomalous QGC's in Eq.~\eqref{eq:HiggsBasisQGCs} that we have omitted for brevity. The relevant Feynman rules may be determined from, for example, Appendix A of~\cite{Dedes:2017zog}.}
\begin{equation}
    d_{Gf},\, d_{Af},\ d_{Zf},\ d_{Wu}, \ d_{Wd},\ d_{We}, \, d_{hGf},\, d_{hAf},\ d_{hZf},\ d_{hWu}, \ d_{hWd},\ d_{hWe} \, .
\end{equation}

The final class of operators of potential interest are four-fermion operators, which may have a large number of CP violating parameters depending on the flavour assumptions. 
These operators are of the same form both in the Higgs basis and the SMEFT Warsaw basis, and have a one-to-one correspondence with HEFT operators. 
For this reason, we will not write these explicitly, though they may be found in~\cite{Dedes:2017zog}.

This general mass basis interactions can be mapped to and from $d=6$ SMEFT parametrisations as in~\cite{Falkowski:2015wza,Azatov:2022kbs}. 
Mapping from a $d=6$ SMEFT operator basis imposes certain relations among the mass eigenbasis couplings, since the above Lagrangian respects fewer symmetries than the full SM gauge symmetry assumed in the SMEFT construction. 
Examples would be relations between some of the gauge-Higgs boson couplings, $\widetilde c_{ww}, \ \widetilde c_{\gamma\gamma}, \ \widetilde c_{z\gamma}, \ \widetilde{c}_{zz}$, as well as a one-to-one mapping between the $HVff$ contact interactions and the corresponding $Vff$ gauge current interactions without the Higgs boson. 

The latter relations are often cited as a reason for neglecting the $HVff$ interactions in LHC studies, given the strong constraints at the level of $10^{-2}\,(10^{-3})$ for the quark (lepton) couplings from EW precision observables at the $Z$-pole measured at LEP~\cite{Efrati:2015eaa}. 
However, it has recently been shown that the LHC data does provide new information on the current-current operators responsible for generating both of these classes of interactions in global fits~\cite{Falkowski:2019hvp,Ellis:2020unq}, such that LHC precision is competing with that of EW precision observables in a global sense. 

More generally, the physical Higgs boson may not come from an $SU(2)_L$ doublet acquiring a vacuum expectation value. 
This is the HEFT ansatz~\cite{Feruglio:1992wf,Bagger:1993zf,Koulovassilopoulos:1993pw,Burgess:1999ha,Grinstein:2007iv,Alonso:2012px,Buchalla:2013rka,Brivio:2013pma,Gavela:2014vra,Cohen:2020xca}, and relaxes many of the relations between Higgs basis operators coming from the SMEFT.
This leaves even more blind directions requiring further observables to break the degeneracy.
More importantly, a number of operators in the HEFT have no counterpart in the SMEFT at $d=6$ or in the Higgs basis as presented above, even restricting to four particle interactions. 
For example, the HEFT has triple gauge couplings with an additional Higgs boson, which would require the following terms in the Higgs basis:
\begin{align}\label{eq:HiggsBasis:htgc}
\begin{split}
 \Delta \mathcal{L}_{\text{htgc}}^\text{CPV} =& \ \frac{h}{v} \bigg[ie \widetilde{\kappa}^{(1)}_\gamma \widetilde{A}_{\mu\nu} W_\mu^+ W_\nu^- +ig_Lc_\theta \widetilde{\kappa}^{(1)}_z \widetilde{Z}_{\mu\nu} W_\mu^+ W_\nu^- + i \frac{e}{m_W^2}\widetilde{\lambda}^{(1)}_\gamma W_{\mu\nu}^+ W_{\nu\rho}^- \widetilde{A}_{\rho\mu} \\&+ i \frac{g_L c_\theta }{m_W^2} \widetilde{\lambda}^{(1)}_z W_{\mu\nu}^+ W_{\nu\rho}^- \widetilde{Z}_{\rho\mu} + \frac{g_s^3}{v^2}\widetilde{c}_{3g}^{(1)} f^{abc} \widetilde{G}_{\mu\nu}^a G_{\nu\rho}^b G_{\rho\mu}^c \, \bigg].
\end{split}
\end{align}
These interactions begin only at $d=8$ in the SMEFT.

\subsection{Experimental parameterizations for CPV\label{subsec:kappas}}

 When the number of EFT operators relevant to a process significantly exceeds the number of available experimental constraints, a complete EFT analysis often becomes impractical in experimental applications. A typical example is a single on-shell $H$ decay channel, where EFT operators can affect not only the production and decay amplitudes but also the total $H$ width, $\Gamma_\mathrm{tot}$, which appears in the propagator denominator.
In these cases, various experimental parametrisations are employed, such as ratios of $\kappa$'s, angle 
parameters~\cite{LHCHiggsCrossSectionWorkingGroup:2012nn,LHCHiggsCrossSectionWorkingGroup:2013rie}, 
or fractional cross sections~\cite{CMS:2021nnc,Anderson:2013afp,Dawson:2013bba}, along with the overall signal strength.
This approach is used to eliminate blind directions by reducing a large set of coupling parameters to a single overall fit parameter, 
such as the signal strength, thus retaining only a small number of parameters that can be effectively constrained in a given channel.

The goal of this Section is to assist theoretical physicists in correctly interpreting experimental 
measurements presented in this notation and applying them effectively in their analyses, as well as informing experimental physicists about what assumptions and information to report to make their results accessible.
We demonstrate that most of the experimental parameterizations mentioned above are equivalent 
to measuring ratios of EFT operators, and in some cases, of linear combinations of these operators.
Assessing the EFT validity of the experimental measurements is beyond the scope of this discussion. 
Such validity is independent of the notation used. If the precision of an experimental result is insufficient 
to be primarily sensitive to the linear terms appearing in interference with the SM process, 
then the resulting constraints are not suitable for a reliable EFT interpretation, regardless of the notation. 
However, the results may still be meaningful for EFT interpretation within a certain validity range which 
must be reported by the experiment, or determined through other means.

To illustrate the above point, it suffices to recognize that the width in the denominator depends on essentially every 
coupling of the $H$, as it represents the sum of partial widths to all possible final states.
For example, the differential distributions for observables $\vec{x}$ can be expressed as
\begin{equation}
\frac{d\sigma(\vec{x})}{d\vec{x}}=
\frac{ 
\left(\sum \gamma^\mathrm{prod}_{jk}c_j c_k\right) 
\left(\sum \gamma^\mathrm{dec}_{lm}c_l c_m\right) }
{\Gamma_\mathrm{tot}(\vec{c}\,)} \,,
\label{eq:crosssection1} 
\end{equation}
where the quantities $\gamma^\mathrm{prod}_{jk}$ correspond to the production cross sections, whereas 
the functions $\gamma^\mathrm{dec}_{lm}$ describe the decay, and either one or both of these may depend on $\vec{x}$.
A single decay channel does not impose constraints on the majority of those couplings $\vec{c}$.
As an example, consider the decays $H \to \tau^+ \tau^-$ or $H \to4\ell$
without tagging a particular production mechanism. 
However, the reasoning applies to any process, either production or decay. 
Although we illustrate this using decay observables $\vec{x}$ below, the approach is equally applicable to production processes.

As a consequence of Eq.~(\ref{eq:crosssection1}), the constraints on individual couplings $c_k$ become degenerate.
However, by simply redefining the parameters as \( c_{m}^\prime \), we can factor out all degenerate couplings 
(enclosed in square brackets), while the remaining terms (in round brackets) describe the differential 
decay distributions as a polynomial in the newly defined parameters:
\begin{equation}
\frac{d\sigma(\vec{x})}{d\vec{x}}=
\left[
\frac{\sum \gamma^\mathrm{prod}_{jk}c_j c_k}{\Gamma_\mathrm{tot}(\vec{c}\,)}
 (c_0+c_1)^2
\right]
 \left(
 \gamma^\mathrm{dec}_{00}(\vec{x})
 + \sum_{1< m} \gamma^\mathrm{dec}_{0m}(\vec{x}) c_m^\prime
 + \sum_{1< m\le l} \gamma^\mathrm{dec}_{lm}(\vec{x}) c_l^\prime c_m^\prime
 \right) \,,
\label{eq:crosssection2} 
\end{equation}
where $\gamma^\mathrm{dec}_{00}(\vec{x})$ describes the SM decay process driven by the coupling $c_0$.
The newly defined parameters $c_m^\prime$ are expressed as ratios of the couplings to the combination \( (c_0 + c_1) \), 
which controls the SM-like tensor structure of the interactions:
\begin{equation}
c_{m}^\prime = c_{m} / (c_0+c_1) \,.
\label{eq:coupleratio} 
\end{equation}

To illustrate the application of Eq.~(\ref{eq:coupleratio}), let us consider the process
$H\rightarrow \bar{f}f$ (or $\bar{f}f\rightarrow H$ as in i.e. $\bar{t} t H$ production), 
where the amplitude may be written as~\cite{Brod:2013cka,Gritsan:2016hjl}
\begin{equation}
    -\frac{m_f}{v}\bar{u}_f (\kappa_f + i \widetilde{\kappa}_f \gamma_5 )u_f
    \label{eq:kappaf} 
\end{equation}
and where $u_f$ is the usual Dirac spinor.
In this case, the $\kappa$'s appearing are directly related to the Higgs basis couplings above via 
\begin{align}
\begin{split}
    {\kappa}_f &= 1+\text{Re}\big(\delta y_f \big) \, ,\\
    \widetilde{\kappa}_f &= \text{Im}\big( \delta y_f\big)\, .
\end{split}
\end{align}
In the notation of Eq.~(\ref{eq:coupleratio}), we have $c_0=1$, $c_1=\text{Re}\big(\delta y_f \big)$, $c_2=\text{Im}\big( \delta y_f\big)$, and 
{$c_2^\prime =\frac{c_2}{c_0+c_1}  = \frac{{\widetilde{\kappa}_f}}{{\kappa_f}}$.

Constraints on the coupling ratios in Eq.~(\ref{eq:coupleratio}) are obtained experimentally from the 
decay differential distributions measured via Eq.~(\ref{eq:crosssection2}). 
Specifically, if constraints on the CP-odd operators are desired, the observables $\vec{x}$ must be CP-sensitive
to isolate the linear terms in the round brackets of Eq.~(\ref{eq:crosssection2}).
The overall signal strength, given by the terms in the square brackets, can either be profiled or, preferably, 
explicitly reported when its interpretation in terms of coupling parameters is well defined.
Ultimately, constraints on coupling ratios and the overall signal strength can be incorporated into global EFT fits, 
provided that their covariance matrix or, preferably, the full likelihood is made available.
In the simplest scenario, a constraint on a single parameter, for example, a CP-sensitive angle,
can be readily incorporated into the EFT fit.

In the discussion above, we have disregarded the operator dimensionality, for the reasons previously outlined.
Although Eqs.~(\ref{eq:crosssection1}) and~(\ref{eq:crosssection2}) include both linear and quadratic terms of dimension-6 
operators without accounting for dimension-8 operators, for EFT validity it is still necessary to demonstrate 
that the contribution of the quadratic terms is negligible and does not affect the results within the reported range. 
In such cases, including or excluding these terms makes no practical difference. 
We retain all terms in Eqs.~(\ref{eq:crosssection1}) and~(\ref{eq:crosssection2}) 
to illustrate the framework and demonstrate one possible approach to assessing validity.
Experimental analyses can still be conducted beyond the EFT validity range, they just cannot be incorporated into global EFT fits.
This can occur if the experimental precision is still limited, or if no CP-sensitive observables can be constructed.
The validity assessment is independent of the notation used.

Next, we demonstrate that constraints on the coupling ratios \( c_{m}^\prime \) in Eq.~(\ref{eq:coupleratio}) are equivalent to setting
bounds on the fractional parameters \( f_{i} \) or the ``mixing angle'' $\alpha_i$ parameters used in experimental analyses.
Following the fermion coupling parameterization in Eq.~(\ref{eq:kappaf}), 
which is relevant, for instance, to the analysis of the decay $H \to \tau^+ \tau^-$, 
three equivalent forms of parameterization can be used:
\begin{align}
r_f &= \frac{\widetilde{\kappa}_f}{\kappa_f} 
\label{eq:threeways1} \\
\alpha^{Hff} &= \tan^{-1}\left( \frac{\widetilde{\kappa}_f}{\kappa_f} \right)
\label{eq:threeways2} \\
f_{\text{CP}}^{Hff} &= \frac{|\widetilde{\kappa}_f|^2}{|\kappa_f|^2 + |\widetilde{\kappa}_f|^2} \, \text{sign}\left( \frac{\widetilde{\kappa}_f}{\kappa_f} \right)
\label{eq:threeways3}
\end{align}
The selection among the three parameterizations is purely historical and does not affect the results. 
When a single ratio is measured, the form given in Eq.~(\ref{eq:threeways2}) is more commonly used, simply for notational convenience.
However, when multiple coupling ratios are measured, defining a separate angle for each becomes impractical. 
As a result, the notation in Eq.~(\ref{eq:threeways3}) is adopted, as illustrated below.
However, we emphasize that the choice of notation conveys no information regarding EFT validity, operator dimensionality, 
whether linear or quadratic terms are used in extracting the results, or any other such aspect.
It is therefore recommended to explicitly state any assumptions made in the analysis when presenting results using this 
or any alternative notation.

The notation in Eq.~(\ref{eq:threeways3}) can be extended, to account for multiple CP-even and CP-odd contributions, as follows
%
\begin{eqnarray}
&& f_{m}=  \frac{|c_{m}^\prime|^2\,\sigma_m/\sigma_1}{1+\sum_{j\ge 2} |c_j^\prime|^2 \, \sigma_j/\sigma_1} 
\text{sign}\left( {c_m^\prime} \right) \,,
\label{eq:fractions1}
\end{eqnarray}
%
where $\sigma_i$ is the effective cross-section of the $H$ decay process corresponding to $c_{i}=1, c_{j \ne i}=0$.
This notation is, for example, well-suited for analyzing the decay $H \to 4\ell$, 
which involve multiple contributions $c_m$, both CP-even and CP-odd, since one can deduce the coupling ratios in terms of the measured fractional contributions:
\begin{eqnarray}
&& c_m^\prime = \sqrt{ \, \frac{\left|f_m\right|}{1 - \sum_{j \ge 2} |f_j|} \frac{\sigma_1}{\sigma_m} \, } \,\text{sign}\left(f_m \right) \,.
\label{eq:fractions2}
\end{eqnarray}}
%
Note that, since linear terms are not included in the denominator of Eq.~(\ref{eq:fractions1}), 
the equation can be easily inverted to obtain Eq.~(\ref{eq:fractions2}).
This establishes the sets \( \{f_m\} \) and \( \{c_m^\prime\} \) as fully equivalent and interchangeable.
This also provides a rationale for omitting linear terms in the notation.

The choice of \( \{f_m\} \) is driven by the intuitive meaning of the parameter as a fractional contribution to the cross section
under a hypothetical scenario of no interference. This choice should not be mistaken for an argument about EFT validity.
It is purely a matter of notation and is unrelated to the specific terms included in the experimental analysis.
In this approach, the specific choice of coupling convention becomes irrelevant. For instance, if the definition of a coupling 
\( c_m \) is rescaled by a constant factor, the corresponding value of \( f_m \) remains unchanged. 
This feature helps avoid potential errors related to coupling conventions, it is sufficient to identify the tensor structure being measured.

Following the above conventions, 
the CP-sensitive parameters $f_{CP}$ were adopted for the benchmark parameter measurements
in feasibility studies, such as in the Snowmass community effort~\cite{Dawson:2013bba,Gritsan:2022php}: 
\begin{equation}
f_{CP}^{HX}\equiv
\frac{\Gamma^{CP\rm\,odd}_{H\to X}
}{\Gamma^{CP\rm\,odd}_{H\to X}
+\Gamma^{CP\rm\,even}_{H\to X}}
\,,
\label{eq:fCP} 
\end{equation}
where a simplified formulation of the partial decay width for 
$H\to X$ was employed for one CP-odd and one CP-even components.

\subsection{SMEFT\label{subsec:SMEFT}}

One of the most common ways of parameterizing new heavy BSM physics is in terms of the SMEFT~\cite{Brivio:2017vri}.
Let us consider the $d=6$ SMEFT in the Warsaw basis~\cite{Grzadkowski:2010es,Dedes:2017zog}.
The conversion to other SMEFT bases may be performed with various tools~\cite{Aebischer:2023nnv,Carmona:2021xtq,Criado:2017khh,Falkowski:2015wza,DasBakshi:2018vni,LopezMiras:2025gar,Fuentes-Martin:2022jrf}.
The relation between the parameters of the SMEFT and those of the Higgs basis are given explicitly in~\cite{Falkowski:2001958,Azatov:2022kbs}.
We ignore any $B$ or $L$ violating operators, since they cannot interfere with the SM and do not contribute to any observable at $\mathcal{O}(1/\Lambda^2)$\,\footnote{With the exception of the dimension-5 Weinberg operator~\cite{Weinberg:1979sa}, which is relevant for neutrino physics.}, where $\Lambda$ is the scale at which the effective operators are generated.

The counting of CP-odd parameters in the SMEFT is given in~\cite{Alonso:2013hga}, and the general set of CP-odd invariants in the Warsaw basis that may contribute at $\mathcal{O}(1/\Lambda^2)$ has been worked out in~\cite{Bonnefoy:2021tbt,Bonnefoy:2023bzx,Darvishi:2024cwe,Sun:2025axx}.
They consist of the Standard Model Jarlskog invariant, the QCD $\theta$ angle, six bosonic CP violating operators, and 699 fermionic invariants.
The CP-odd bosonic operators are given by
\begin{align}\label{eq:SMEFTbosonic}
\begin{split}
    \mathcal{L}^\text{CPV}_\text{bosonic} = \frac{1}{\Lambda^2}\bigg[ &C_{H\widetilde{W}}H^\dagger H \widetilde{W}^I_{\mu\nu} W^{I,\mu\nu}+ C_{H\widetilde{B}}H^\dagger H \widetilde{B}_{\mu\nu} B^{\mu\nu} +C_{H\widetilde{W}B}H^\dagger \tau^I H \widetilde{W}^I_{\mu\nu}B^{\mu\nu}\\
    &+ C_{H\widetilde{G}} H^\dagger H \widetilde{G}^{A}_{\mu\nu} G^{A,\mu\nu}+C_{\widetilde{G}} f^{ABC}\widetilde{G}^{A,\nu}_\mu G_\nu^{B,\rho} G_\rho^{C,\mu} + C_{\widetilde{W}} \epsilon^{IJK}\widetilde{W}^{I,\nu}_\mu W_\nu^{J,\rho} W_\rho^{K,\mu}\bigg] \ ,
\end{split}
\end{align}
where $G^a_{\mu\nu}$, $W^i_{\mu\nu}$ 
and $B_{\mu\nu}$ are the $\mathrm{SU}(3)_C \times \mathrm{SU}(2)_L \times \mathrm{U}(1)_Y$ field strengths, $\tilde{X}_{\mu\nu}\equiv \frac{1}{2} \epsilon_{\mu\nu\rho\sigma} {X}^{\rho\sigma}$, with $\epsilon^{0123}=1$, their associated dual fields and $H$ is the Higgs doublet.
In the two-fermion sector, the dipole operators have coefficients that are complex $3 \times 3$ matrices:
\begin{align}\label{eq:SMEFTdipoles}
\begin{split}
    \mathcal{L}_\text{dipole} = \frac{1}{\Lambda^2}\bigg[&C_{eW}[ij](\bar{\ell}_{L,i}\sigma^{\mu\nu}e_{R,j})\tau^I H W_{\mu\nu}^I + C_{eB}[ij](\bar{\ell}_{L,i}\sigma^{\mu\nu}e_{R,j})H B_{\mu\nu} \\
    &+ C_{uG}[ij](\bar{q}_{L,i}\sigma^{\mu\nu} T^au_{R,j}) \widetilde{H} G_{\mu\nu}^A + C_{uW}[ij](\bar{q}_{L,i}\sigma^{\mu\nu} u_{R,j}) \tau^I \widetilde{H} W^I_{\mu\nu} \\
    & + C_{uB}[ij](\bar{q}_{L,i}\sigma^{\mu\nu} u_{R,j}) \widetilde{H} B_{\mu\nu} +
    C_{dG}[ij](\bar{q}_{L,i}\sigma^{\mu\nu} T^a d_{R,j}) H G^A_{\mu\nu} \\
    &+ C_{dW}[ij](\bar{q}_{L,i}\sigma^{\mu\nu} d_{R,j}) \tau^I H W^I_{\mu\nu} + C_{dB}[ij](\bar{q}_{L,i}\sigma^{\mu\nu} d_{R,j}) H B_{\mu\nu}
    \bigg]  \ .
\end{split}
\end{align}
Similarly, there are the Yukawa-like operators 
\begin{align}\label{eq:SMEFTYukawa}
\begin{split}
  \mathcal{L}_\text{Yukawa} = \frac{1}{\Lambda^2}\bigg[&C_{eH}[ij] (H^\dagger H)(\bar{\ell}_{L,i}e_{R,j} H) +C_{uH}[ij] (H^\dagger H)(\bar{q}_{L,i}u_{R,j} \widetilde{H}) \\ &+C_{dH}[ij]( H^\dagger H)(\bar{q}_{L,i}d_{R,j} H) \bigg] \ ,
\end{split}
\end{align}
with coefficients that are also complex $3 \times 3$ matrices.
Finally, there are the gauge-fermion interactions
\begin{align}\label{eq:SMEFTgauge}
\begin{split}
 \mathcal{L}_{\psi^2H^2D} = & \ \frac{1}{\Lambda^2}\bigg[C_{H\ell}^{(1)}[ij] (H^\dagger i \overleftrightarrow{D_\mu} H ) (\bar{\ell}_{L,i} \gamma^\mu \ell_{L,j}) +C_{H\ell}^{(3)}[ij] (H^\dagger i \overleftrightarrow{D^I_\mu} H ) (\bar{\ell}_{L,i} \tau^I \gamma^\mu \ell_{L,j}) \\ 
 & \qquad +C_{He}[ij] (H^\dagger i \overleftrightarrow{D_\mu} H )(\bar{e}_{R,i} \gamma^\mu e_{R,j}) +C_{Hq}^{(1)}[ij] (H^\dagger i \overleftrightarrow{D_\mu} H ) (\bar{q}_{L,i} \gamma^\mu q_{L,j})\\
 & \qquad  +C_{Hq}^{(3)}[ij] (H^\dagger i \overleftrightarrow{D^I_\mu} H ) (\bar{q}_{L,i} \tau^I \gamma^\mu q_{L,j}) +C_{Hu}[ij] (H^\dagger i \overleftrightarrow{D_\mu} H )(\bar{u}_{R,i} \gamma^\mu u_{R,j})  \\
 & \qquad +C_{Hd}[ij] (H^\dagger i \overleftrightarrow{D_\mu} H )(\bar{d}_{R,i} \gamma^\mu d_{R,j})+C_{Hud}[ij](\widetilde{H}^\dagger i D_\mu H) (\bar{u}_{R,i} \gamma^\mu d_{R,j})\bigg] \ ,
\end{split}
\end{align}
where $H^\dagger i\overleftrightarrow{ D_\mu}H = iH^\dagger (D_\mu - \overleftarrow{D_\mu})H$ and $H^\dagger i\overleftrightarrow{ D^I_\mu}H = iH^\dagger (\tau^ID_\mu - \overleftarrow{D_\mu}\tau^I)H$, which lead to changes in the $V\bar{f}f$ couplings after EWSB.
Here the coefficient $C_{Hud}[ij]$ is a general complex $3 \times 3$ matrix, while the remainder are Hermitian and may only be CP violating when there are multiple generations.

Finally, all four-fermion operators in the SMEFT may contribute to CP violation. 
The number of CP violating parameters in the two-fermion and four-fermion operators greatly depends on the flavour assumptions imposed.
The four-fermion operators are generally less important for Higgs boson phenomenology compared to two-fermion and bosonic operators, and so we will not write them explicitly, see however\,\cite{Dedes:2017zog} for the full set.  We have written all of the above interactions in the gauge basis, but in practice the coefficients that enter observables come after rotating the fermion fields into the mass basis. 
We follow the conventions in~\cite{Dedes:2017zog}, which absorbs some rotation matrices into the SMEFT coefficients so that the Feynman rules depend only on SMEFT matrices and CKM/PMNS matrices.

\subsubsection{Flavour assumptions}

In full generality, all two-fermion operators and four-fermion operators have coefficients which are complex matrices involving many CP violating phases. 
However, since the 2499 Baryon-number conserving operators~\cite{Alonso:2013hga,Alonso:2014zka,Henning:2015alf} present in the SMEFT with no flavour assumptions are too many to simultaneously constrain them in full generality, some additional flavour symmetries are often imposed~\cite{Faroughy:2020ina,Brivio:2020onw,Degrande:2021zpv}. 
The most common choices are those listed in~\cite{Brivio:2020onw} and~\cite{Bonnefoy:2021tbt,Bonnefoy:2023bzx,Darvishi:2024cwe,Sun:2025axx} in the 
context of CP violation.

In the absence of the Yukawa couplings, the SM has a $U(3)^5$ flavour symmetry. 
If this full symmetry is imposed, then many two-fermion and four-fermion coefficients become CP-even and diagonal.
In the Minimal Flavour Violation (MFV) hypothesis~\cite{Chivukula:1987py,DAmbrosio:2002vsn},\footnote{Sometimes the MFV ansatz additionally assumes that the SM Yukawa matrices are the only source of CP violation~\cite{DAmbrosio:2002vsn,Brivio:2020onw}, which sets all new CP violating operators to zero.} the SM Yukawa matrices are considered spurions of this $U(3)^5$, and working beyond zeroth order induces a handful of new CP violating operators. See~\cite{Faroughy:2020ina} for the counting, which depends on the order in each of $Y_e$, $Y_u$, and $Y_d$. 
Going beyond MFV and $U(3)^5$, any subgroup of the $U(3)^5$ may be imposed in principle. 
In practice, at minimum a $U(2)^3$ symmetry is typically imposed on the quark sector to avoid stringent flavour constraints on the first two generations, and either a $U(1)^3$ or $U(3)^2$ symmetry is imposed in the lepton sector. 

Which CP violating operators vanish after imposing a flavour symmetry can be determined by whether the corresponding coefficient matrix is hermitian or complex -- see Appendix A of~\cite{Alonso:2013hga}. 
For example, the coefficients of the dipole operators in Eq.~\eqref{eq:SMEFTdipoles} are general complex $3 \times 3$ matrices, and so they may still have a CP violating complex entry even assuming a $U(3)^5$ symmetry.
On the other hand, the coefficients of the operators in Eq.~\eqref{eq:SMEFTgauge} (with the exception of $C_{Hud}$), are hermitian, and become strictly CP preserving in the $U(3)^5$ limit.
Note that this is also the case if a $U(2)^3$ symmetry is applied, as is common when singling out third generation quarks.
These conclusions hold regardless of whether one is working in the SMEFT, the HEFT, or the Higgs basis.

\subsection{HEFT\label{subsec:HEFT}}

The most general possible EFT using the SM degrees of freedom is the Higgs Effective Field Theory (HEFT).
The idea behind HEFT relies on treating the Higgs boson as a singlet of the theory instead of as part of $SU(2)_L$ doublet, see~\cite{Feruglio:1992wf,Bagger:1993zf,Koulovassilopoulos:1993pw,Burgess:1999ha,Grinstein:2007iv,Alonso:2012px,Buchalla:2013rka,Brivio:2013pma,Gavela:2014vra,Cohen:2020xca} among others.
For the discussion we will follow the conventions in~\cite{Brivio:2016fzo}.
The most important feature of the HEFT compared to the SMEFT for Higgs boson physics is that  operators with one additional Higgs boson insertion are independent, which can have significant impact on phenomenology. 
In~\cite{Brivio:2016fzo} they assume flavour diagonality, which removes some sources of CP violation.
Extension to nontrivial flavour structures in the HEFT is largely the same as in the SMEFT, though comprehensive HEFT studies of CP invariants analogous to~\cite{Bonnefoy:2021tbt,Bonnefoy:2023bzx,Darvishi:2024cwe,Sun:2025axx} do not yet exist to our knowledge.
The building blocks of the HEFT are the $U(1)_Y\times SU(2)_L \times SU(3)_c$ gauge field strengths $B_{\mu\nu}$, $W_{\mu\nu}$, $G_{\mu\nu}$, the scalar chiral fields $\mathbf{V_\mu}$ and $\mathbf{T}$, and the functions $\mathcal{F}(h)$, where 
\begin{align}
    \mathbf{V}_\mu &\equiv(\mathbf{D_\mu U})\mathbf{U}^\dagger\\
    \mathbf{T}&\equiv\mathbf{U}\sigma_3 \mathbf{U}^\dagger\\
    \mathcal{F}_i(h)&\equiv1+2a_i\dfrac{h}{v}+b_i \dfrac{h^2}{v^2},
\end{align}
where $\mathbf{U}(x) \equiv e^{i \sigma_a \pi_a(x)/v}$ with $\pi_a$ the $SU(2)_L$ Goldstone bosons and 
\begin{equation}
\mathbf{D_\mu U}(x)\equiv \partial_\mu \mathbf{U}(x)+ i g W_\mu (x) \mathbf{U}(x)-\dfrac{i g^\prime}{2}B_\mu(x)\mathbf{U}(x)\sigma_3 .
\end{equation}
For simplicity, we will work in the unitary gauge, where $\mathbf{U}(x) = \mathds{1}$ and $\mathbf{T}=\sigma_3$.

The CP-odd gauge and gauge-Higgs boson operators were presented in~\cite{Gavela:2014vra,Hierro:2015nna}.
Including also fermionic operators, the full HEFT Lagrangian can be found in~\cite{Brivio:2016fzo}, which included a number of equation of motion simplifications to obtain a complete basis.
Following ~\cite{Brivio:2016fzo}, we will work with the leading order (LO) and next to leading order (NLO) terms of the full HEFT Lagrangian. 
Grouping the SM fermions in $SU(2)_{L,R}$ doublets
\begin{equation}
    Q_L=\binom{U_L}{D_L}, \quad Q_R=\binom{U_R}{D_R}, \quad L_L=\binom{\nu_L}{E_L}, \quad L_R=\binom{0}{E_R},
\end{equation}
one can write the HEFT Lagrangian as a sum of two terms,
\begin{equation}
    \mathcal{L}_{\mathrm{HEFT}} \equiv \mathcal{L}_0+\Delta \mathcal{L},
\end{equation}
where the first term stands for LO operators and the second one collects NLO operators.

The LO Lagrangian includes the kinetic terms for all the particles in the spectrum, the Yukawa interactions and the scalar potential, namely
\begin{align}\label{eq:HEFTLO}
\begin{split}
\mathcal{L}_0= & -\frac{1}{4} G_{\mu \nu}^\alpha G^{\alpha, \mu \nu}-\frac{1}{4} W_{\mu \nu}^a W^{a, \mu \nu}-\frac{1}{4} B_{\mu \nu} B^{\mu \nu}
\\
& +\frac{1}{2} \partial_\mu h \partial^\mu h-\frac{v^2}{4} \operatorname{Tr}\left(\mathbf{V}_\mu \mathbf{V}^\mu\right) \mathcal{F}_C(h)-V(h) 
\\
& +i \bar{Q}_L \slashed{D} Q_L+i \bar{Q}_R \slashed D Q_R+i \bar{L}_L \slashed D L_L+i \bar{L}_R \slashed D L_R \\
& -\frac{v}{\sqrt{2}}\left(\bar{Q}_L \mathbf{U} \mathcal{Y}_Q(h) Q_R+\text { h.c. }\right)-\frac{v}{\sqrt{2}}\left(\bar{L}_L \mathbf{U} \mathcal{Y}_L(h) L_R+\text { h.c. }\right) 
\\
& -\frac{g_s^2}{16 \pi^2} \lambda_s G_{\mu \nu}^\alpha \tilde{G}^{\alpha, \mu \nu},
\end{split}
\end{align}
where the last line is the CP violating QCD theta angle. 
Here the function multiplying the Goldstone kinetic term is given by
\begin{equation}
    \mathcal{F}_C(h) = 1+2(1-\Delta a_C)(h/v) + (1-\Delta b_C) (h/v)^2 + \ldots
\end{equation}
where $\Delta a_C$ and $\Delta b_C$ are NLO sized. 
These are strictly CP-even, and will not be of interest to us here.
The Yukawa matrices are given by
\begin{equation}
    \mathcal{Y}_Q(h) = \text{diag}\left( \sum^{\infty}_{n=0}  \mathcal{Y}_u^{(n)} \frac{h^n}{v^n}, \sum^{\infty}_{n=0}  \mathcal{Y}_d^{(n)} \frac{h^n}{v^n}\right)\, , \qquad \mathcal{Y}_L(h) = \text{diag}\left(0, \sum^{\infty}_{n=0}  \mathcal{Y}_\ell^{(n)} \frac{h^n}{v^n}\right) \, ,
\end{equation}
where $\mathcal{Y}^{(0)}_f$ gives the fermion masses and CKM matrix after diagonalization, and $\mathcal{Y}^{(1)}_f = \mathcal{Y}^{(0)}_f + \Delta \mathcal{Y}_f^{(1)}$ gives the $H\bar{f} f$ Yukawa couplings, where $\Delta \mathcal{Y}_f^{(1)}$ is taken to be NLO sized.
All of $\mathcal{Y}^{(n)}_f$ may be non-aligned and have CP violating phases in general, though they vanish in the $U(3)^5$ limit.

At NLO, we are interested in those operators that are CP violating, which we divide depending on their particle content as bosonic and fermionic operators in the subsections below.
Starting with the CP-odd bosonic sector, the Lagrangian contains 16 operators
\begin{equation}
\label{eq:HEFT-bos}
   \Delta \mathcal{L}_{\text {bos }}^{\text{CPV}}=\sum_{j} \widetilde{c}_{S_j}\mathcal{S}_j,\quad j=\{2D,\, \widetilde{B},\, \widetilde{W},\,\widetilde{G},\,1-9,\,15,\,\widetilde{W}WW,\widetilde{G}GG\}\ .
\end{equation}
The expressions for the operators are given by
$$
\mathcal{S}_{2 D}(h) \equiv i \frac{v^2}{4} \operatorname{Tr}\left(\mathbf{T} \mathbf{V}_\mu\right) \partial^\mu \mathcal{F}_{2 D},
$$
$$
\begin{aligned}
\mathcal{S}_{\widetilde{W} W W}(h) & =\frac{4 \pi \varepsilon_{a b c}}{\Lambda^2} \widetilde{W}_\mu^{a \nu} W_\nu^{b \rho} W_\rho^{c \mu} \mathcal{F}_{\widetilde{W} W W}, \\
\mathcal{S}_{\widetilde{G} G G}(h) & =\frac{4 \pi f_{\alpha \beta \gamma}}{\Lambda^2} \widetilde{G}_\mu^{\alpha \nu} G_\nu^{\beta \rho} G_\rho^{\gamma \mu} \mathcal{F}_{\widetilde{G} G G},
\end{aligned}
$$
\begin{align}
\begin{array}{ll}
\vspace{0.1cm}
\mathcal{S}_{\widetilde{B}}(h) \equiv-B^{\mu \nu} \widetilde{B}_{\mu \nu} \mathcal{F}_{\widetilde{B}} & \mathcal{S}_{\widetilde{W}}(h) \equiv-\operatorname{Tr}\left(W^{\mu \nu} \widetilde{W}_{\mu \nu}\right) \mathcal{F}_{\widetilde{W}} \\[4pt]
\vspace{0.1cm}
\mathcal{S}_{\widetilde{G}}(h) \equiv-G^{a \mu \nu} \widetilde{G}_{\mu \nu}^a \mathcal{F}_{\widetilde{G}} & \mathcal{S}_1(h) \equiv \widetilde{B}^{\mu \nu} \operatorname{Tr}\left(\mathbf{T} W_{\mu \nu}\right) \mathcal{F}_1 \\[4pt]
\vspace{0.1cm}
\mathcal{S}_2(h) \equiv \frac{i}{4 \pi} \widetilde{B}^{\mu \nu} \operatorname{Tr}\left(\mathbf{T} \mathbf{V}_\mu\right) \partial_\nu \mathcal{F}_2 & \mathcal{S}_3(h) \equiv \frac{i}{4 \pi} \operatorname{Tr}\left(\widetilde{W}^{\mu \nu} \mathbf{V}_\mu\right) \partial_\nu \mathcal{F}_3 \\[4pt]
\vspace{0.1cm}
\mathcal{S}_4(h) \equiv \frac{1}{4 \pi} \operatorname{Tr}\left(W^{\mu \nu} \mathbf{V}_\mu\right) \operatorname{Tr}\left(\mathbf{T} \mathbf{V}_\nu\right) \mathcal{F}_4 & \mathcal{S}_5(h) \equiv \frac{i}{(4 \pi)^2} \operatorname{Tr}\left(\mathbf{V}^\mu \mathbf{V}^\nu\right) \operatorname{Tr}\left(\mathbf{T} \mathbf{V}_\mu\right) \partial_\nu \mathcal{F}_5 \\[4pt]
\vspace{0.1cm}
\mathcal{S}_6(h) \equiv \frac{i}{(4 \pi)^2} \operatorname{Tr}\left(\mathbf{V}^\mu \mathbf{V}_\mu\right) \operatorname{Tr}\left(\mathbf{T} \mathbf{V}^\nu\right) \partial_\nu \mathcal{F}_6 & \mathcal{S}_7(h) \equiv\operatorname{Tr}\left(\mathbf{T}\left[W^{\mu \nu}, \mathbf{V}_\mu\right]\right) \partial_\nu \mathcal{F}_7\\[4pt]
\vspace{0.1cm}
\mathcal{S}_8(h) \equiv \operatorname{Tr}\left(\mathbf{T} \widetilde{W}^{\mu \nu}\right) \operatorname{Tr}\left(\mathbf{T} W_{\mu \nu}\right) \mathcal{F}_8 &  \mathcal{S}_9(h) \equiv \frac{i}{4 \pi} \operatorname{Tr}\left(\mathbf{T} \widetilde{W}^{\mu \nu}\right) \operatorname{Tr}\left(\mathbf{T} \mathbf{V}_\mu\right) \partial_\nu \mathcal{F}_9\\[4pt]
\vspace{0.1cm}
\mathcal{S}_{15}(h) \equiv \frac{i}{(4 \pi)^2} \operatorname{Tr}\left(\mathbf{T} \mathbf{V}^\mu\right)\left(\operatorname{Tr}\left(\mathbf{T} \mathbf{V}^\nu\right)\right)^2 \partial_\mu \mathcal{F}_{15} \ , &
\end{array}
\end{align}
where $\Lambda$ is the UV scale at which the effective operators are generated. The operators with single fermionic currents and up to two derivatives that will be used in our analysis are contained in the following Lagrangian
\begin{equation}
\label{eq:HEFT-2f}
\begin{aligned}
\Delta \mathcal{L}_{2 F}^\text{CPV}= & \sum_{j=1}^{8} n_{j}^{\mathcal{Q}} \mathcal{N}_{j}^{\mathcal{Q}}+\sum_{j=9}^{28} \frac{i}{\Lambda} \tilde{n}_{j}^{\mathcal{Q}} \mathcal{N}_{j}^{\mathcal{Q}}+\sum_{j=29}^{36} \frac{4 \pi i}{\Lambda} \tilde{n}_{j}^{\mathcal{Q}} \mathcal{N}_{j}^{\mathcal{Q}} \\
& + \sum_{j=1}^{2}n_{j}^{\ell} \mathcal{N}_{j}^{\ell}+\sum_{j=3}^{11} \frac{i}{\Lambda} \tilde{n}_{j}^{\ell} \mathcal{N}_{j}^{\ell}+\sum_{j=12}^{14} \frac{4 \pi i}{\Lambda} \tilde{n}_{j}^{\ell} \mathcal{N}_{j}^{\ell}+\text { h.c. } \ ,
\end{aligned}
\end{equation}
 where the intrinsically CP violating operators that have an imaginary coefficient are defined below
\begin{equation}\label{eq:heft_ferm_current_1}
\begin{array}{l l}
\vspace{0.1cm}
\mathcal{N}_{9}^{\mathcal{Q}}(h) \equiv \overline{Q}_{L} \mathbf{U} Q_{R} \partial_{\mu} \mathcal{F} \partial^{\mu} \mathcal{F}^{\prime} & \mathcal{N}_{10}^{\mathcal{Q}}(h) \equiv \overline{Q}_{L} \mathbf{T} \mathbf{U} Q_{R} \partial_{\mu} \mathcal{F} \partial^{\mu} \mathcal{F}^{\prime} \\[4pt] 
\vspace{0.1cm}
\mathcal{N}_{11}^{\mathcal{Q}}(h) \equiv \overline{Q}_{L} \mathbf{V}_{\mu} \mathbf{U} Q_{R} \partial^{\mu} \mathcal{F} & \mathcal{N}_{12}^{\mathcal{Q}}(h) \equiv \overline{Q}_{L}\left\{\mathbf{V}_{\mu}, \mathbf{T}\right\} \mathbf{U} Q_{R} \partial^{\mu} \mathcal{F} \\[4pt]
\vspace{0.1cm}
\mathcal{N}_{13}^{\mathcal{Q}}(h) \equiv \overline{Q}_{L}\left[\mathbf{V}_{\mu}, \mathbf{T}\right] \mathbf{U} Q_{R} \partial^{\mu} \mathcal{F} & \mathcal{N}_{14}^{\mathcal{Q}}(h) \equiv \overline{Q}_{L} \mathbf{T} \mathbf{V}_{\mu} \mathbf{T} \mathbf{U} Q_{R} \partial^{\mu} \mathcal{F} \\[4pt]
\vspace{0.1cm}
\mathcal{N}_{15}^{\mathcal{Q}}(h) \equiv \overline{Q}_{L} \mathbf{V}_{\mu} \mathbf{V}^{\mu} \mathbf{U} Q_{R} \mathcal{F} & \mathcal{N}_{16}^{\mathcal{Q}}(h) \equiv \overline{Q}_{L} \mathbf{V}_{\mu} \mathbf{V}^{\mu} \mathbf{T} \mathbf{U} Q_{R} \mathcal{F} \\[4pt]
\vspace{0.1cm}
\mathcal{N}_{17}^{\mathcal{Q}}(h) \equiv \overline{Q}_{L} \mathbf{T} \mathbf{V}_{\mu} \mathbf{T} \mathbf{V}^{\mu} \mathbf{U} Q_{R} \mathcal{F} & \mathcal{N}_{18}^{\mathcal{Q}}(h) \equiv \overline{Q}_{L} \mathbf{T} \mathbf{V}_{\mu} \mathbf{T} \mathbf{V}^{\mu} \mathbf{T} \mathbf{U} Q_{R} \mathcal{F} \\[4pt]
\vspace{0.1cm}
\mathcal{N}_{19}^{\mathcal{Q}}(h) \equiv \overline{Q}_{L} \mathbf{V}_{\mu} \mathbf{T} \mathbf{V}^{\mu} \mathbf{U} Q_{R} \mathcal{F} & \mathcal{N}_{20}^{\mathcal{Q}}(h) \equiv \overline{Q}_{L} \mathbf{V}_{\mu} \mathbf{T} \mathbf{V}^{\mu} \mathbf{T} \mathbf{U} Q_{R} \mathcal{F}\\[4pt]
\vspace{0.1cm}
\mathcal{N}_{21}^{\mathcal{Q}}(h) \equiv \bar{Q}_L \sigma^{\mu \nu} \mathbf{V}_\mu \mathbf{U} Q_R \partial_\nu \mathcal{F} & \mathcal{N}_{22}^{\mathcal{Q}}(h) \equiv \bar{Q}_L \sigma^{\mu \nu}\left[\mathbf{V}_\mu, \mathbf{T}\right] \mathbf{U} Q_R \partial_\nu \mathcal{F} \\[4pt]
\vspace{0.1cm}
\mathcal{N}_{23}^{\mathcal{Q}}(h) \equiv \bar{Q}_L \sigma^{\mu \nu}\left\{\mathbf{V}_\mu, \mathbf{T}\right\} \mathbf{U} Q_R \partial_\nu \mathcal{F} & \mathcal{N}_{24}^{\mathcal{Q}}(h) \equiv \bar{Q}_L \sigma^{\mu \nu} \mathbf{T} \mathbf{V}_\mu \mathbf{T} \mathbf{U} Q_R \partial_\nu \mathcal{F} \\[4pt]
\vspace{0.1cm}
\mathcal{N}_{25}^{\mathcal{Q}}(h) \equiv \bar{Q}_L \sigma^{\mu \nu} \mathbf{V}_\mu \mathbf{T} \mathbf{V}_\nu \mathbf{U} Q_R \mathcal{F} & \mathcal{N}_{26}^{\mathcal{Q}}(h) \equiv \bar{Q}_L \sigma^{\mu \nu} \mathbf{V}_\mu \mathbf{T} \mathbf{V}_\nu \mathbf{T} \mathbf{U} Q_R \mathcal{F} \\[4pt]
\vspace{0.1cm}
\mathcal{N}_{27}^{\mathcal{Q}}(h) \equiv \bar{Q}_L \sigma^{\mu \nu}\left[\mathbf{V}_\mu, \mathbf{V}_\nu\right] \mathbf{U} Q_R \mathcal{F} & \mathcal{N}_{28}^{\mathcal{Q}}(h) \equiv \bar{Q}_L \sigma^{\mu \nu}\left[\mathbf{V}_\mu, \mathbf{V}_\nu\right] \mathbf{T} \mathbf{U} Q_R \mathcal{F} \\[4pt]
\vspace{0.1cm}
\mathcal{N}_{29}^{\mathcal{Q}}(h) \equiv i g^{\prime} \bar{Q}_L \sigma^{\mu \nu} \mathbf{U} Q_R B_{\mu \nu} \mathcal{F} & \mathcal{N}_{30}^{\mathcal{Q}}(h) \equiv i g^{\prime} \bar{Q}_L \sigma^{\mu \nu} \mathbf{T} \mathbf{U} Q_R B_{\mu \nu} \mathcal{F} \\[4pt]
\vspace{0.1cm}
\mathcal{N}_{31}^{\mathcal{Q}}(h) \equiv i g_s \bar{Q}_L \sigma^{\mu \nu} G_{\mu \nu} \mathbf{U} Q_R \mathcal{F} & \mathcal{N}_{32}^{\mathcal{Q}}(h) \equiv i g_s \bar{Q}_L \sigma^{\mu \nu} G_{\mu \nu} \mathbf{T} \mathbf{U} Q_R \mathcal{F} \\[4pt]
\vspace{0.1cm}
\mathcal{N}_{33}^{\mathcal{Q}}(h) \equiv i g \bar{Q}_L \sigma^{\mu \nu} W_{\mu \nu} \mathbf{U} Q_R \mathcal{F} & \mathcal{N}_{34}^{\mathcal{Q}}(h) \equiv i g \bar{Q}_L \sigma^{\mu \nu}\left\{W_{\mu \nu}, \mathbf{T}\right\} \mathbf{U} Q_R \mathcal{F} \\[4pt]
\vspace{0.1cm}
\mathcal{N}_{35}^{\mathcal{Q}}(h) \equiv i g \bar{Q}_L \sigma^{\mu \nu}\left[W_{\mu \nu}, \mathbf{T}\right] \mathbf{U} Q_R \mathcal{F} & \mathcal{N}_{36}^{\mathcal{Q}}(h) \equiv i g \bar{Q}_L \sigma^{\mu \nu} \mathbf{T} W_{\mu \nu} \mathbf{T} \mathbf{U} Q_R \mathcal{F}  \ ,
\end{array}
\end{equation}
and
\begin{equation}\label{eq:heft_ferm_current_2}
\begin{array}{ll}
\vspace{0.1cm}
\mathcal{N}_3^{\ell}(h) \equiv \bar{L}_L \mathbf{U} L_R \partial_\mu \mathcal{F} \partial^\mu \mathcal{F}^{\prime} & \mathcal{N}_4^{\ell}(h) \equiv \bar{L}_L\left\{\mathbf{V}_\mu, \mathbf{T}\right\} \mathbf{U} L_R \partial^\mu \mathcal{F} \\[4pt]
\vspace{0.1cm}
\mathcal{N}_5^{\ell}(h) \equiv \bar{L}_L\left[\mathbf{V}_\mu, \mathbf{T}\right] \mathbf{U} L_R \partial^\mu \mathcal{F} & \mathcal{N}_6^{\ell}(h) \equiv \bar{L}_L \mathbf{V}_\mu \mathbf{V}^\mu \mathbf{U} L_R \mathcal{F} \\[4pt]
\vspace{0.1cm}
\mathcal{N}_7^{\ell}(h) \equiv \bar{L}_L \mathbf{T} \mathbf{V}_\mu \mathbf{T} \mathbf{V}^\mu \mathbf{U} L_R \mathcal{F} & \mathcal{N}_8^{\ell}(h) \equiv \bar{L}_L \sigma^{\mu \nu}\left[\mathbf{V}_\mu, \mathbf{T}\right] \mathbf{U} L_R \partial_\nu \mathcal{F} \\[4pt]
\vspace{0.1cm}
\mathcal{N}_9^{\ell}(h) \equiv \bar{L}_L \sigma^{\mu \nu}\left\{\mathbf{V}_\mu, \mathbf{T}\right\} \mathbf{U} L_R \partial_\nu \mathcal{F} & \mathcal{N}_{10}^{\ell}(h) \equiv \bar{L}_L \sigma^{\mu \nu} \mathbf{V}_\mu \mathbf{T} \mathbf{V}_\nu \mathbf{U} L_R \mathcal{F} \\[4pt]
\vspace{0.1cm}
\mathcal{N}_{11}^{\ell}(h) \equiv \bar{L}_L \sigma^{\mu \nu}\left[\mathbf{V}_\mu, \mathbf{V}_\nu\right] \mathbf{U} L_R \mathcal{F} & \mathcal{N}_{12}^{\ell}(h) \equiv i g^{\prime} \bar{L}_L \sigma^{\mu \nu} \mathbf{U} L_R B_{\mu \nu} \mathcal{F} \\[4pt]
\vspace{0.1cm}
\mathcal{N}_{13}^{\ell}(h) \equiv i g \bar{L}_L \sigma^{\mu \nu} W_{\mu \nu} \mathbf{U} L_R \mathcal{F} & \mathcal{N}_{14}^{\ell}(h) \equiv i g \bar{L}_L \sigma^{\mu \nu}\left[W_{\mu \nu}, \mathbf{T}\right] \mathbf{U} L_R \mathcal{F}.
\end{array}
\end{equation}
On top of these, there are three additional operators, which are intrinsically CP violating and are
\begin{equation}
    \mathcal{N}_3^Q\equiv\overline{Q_L} \gamma_\mu \left[\mathbf{V}^\mu,\mathbf{T}\right] Q_L \mathcal{F} \quad \mathcal{N}_4^Q\equiv \overline{Q_R} \gamma_\mu \mathbf{U}^\dagger\left[\mathbf{V}^\mu,\mathbf{T}\right] \mathbf{U} Q_R \mathcal{F}\quad \mathcal{N}_1^\ell\equiv \overline{L_L} \gamma_\mu \left[\mathbf{V}^\mu,\mathbf{T}\right] L_L \mathcal{F} \ .
\end{equation}
Additionally, there are some operators that were considered to have strictly real coefficients in~\cite{Brivio:2016fzo} but may generically have CP violating imaginary flavour off-diagonal elements in general, which we will include:
\begin{equation}
\begin{array}{ll}
\vspace{0.1cm}
\mathcal{N}_1^{\mathcal{Q}}(h) \equiv i \bar{Q}_L \gamma_\mu \mathbf{V}^\mu Q_L \mathcal{F} &\mathcal{N}_2^{\mathcal{Q}}(h)  \equiv i \bar{Q}_R \gamma_\mu \mathbf{U}^{\dagger} \mathbf{V}^\mu \mathbf{U} Q_R \mathcal{F}\\
\vspace{0.1cm}
\mathcal{N}_5^{\mathcal{Q}}(h) \equiv i \bar{Q}_L \gamma_\mu\left\{\mathbf{V}^\mu, \mathbf{T}\right\} Q_L \mathcal{F} &\mathcal{N}_6^{\mathcal{Q}}(h)  \equiv i \bar{Q}_R \gamma_\mu \mathbf{U}^{\dagger}\left\{\mathbf{V}^\mu, \mathbf{T}\right\} \mathbf{U} Q_R \mathcal{F}\\
\vspace{0.1cm}
\mathcal{N}_7^{\mathcal{Q}}(h) \equiv i \bar{Q}_L \gamma_\mu \mathbf{T} \mathbf{V}^\mu \mathbf{T} Q_L \mathcal{F} &\mathcal{N}_8^{\mathcal{Q}}(h) \equiv i \bar{Q}_R \gamma_\mu \mathbf{U}^{\dagger} \mathbf{T} \mathbf{V}^\mu \mathbf{T} \mathbf{U} Q_R \mathcal{F},
\end{array}
\end{equation}
for quarks and
\begin{equation}
\mathcal{N}_2^\ell\equiv i\overline{L_R} \gamma_\mu \mathbf{U}^\dagger\left\{\mathbf{V}^\mu,\mathbf{T}\right\}\mathbf{U} L_R \mathcal{F} \ ,
\end{equation}
for leptons.
In full flavour generality, there are also the operators (Eq. (2.21) of~\cite{Brivio:2016fzo}):
\begin{align}
    \mathcal{N}_{\text{red},1}^{\ell}\equiv i\bar{L}_{L_i} \gamma_\mu \mathbf{V}^\mu L_{L_j}\mathcal{F}\, , \quad \mathcal{N}_{\text{red},2}^{\ell}\equiv i\bar{L}_{L_i}\{\mathbf{T},\mathbf{V}^\mu\}L_{L_j} \mathcal{F}\, , \quad \mathcal{N}_{\text{red},3}^{\ell} \equiv i\bar{L}_{L_i} \gamma_\mu \mathbf{T} \mathbf{V}^\mu \mathbf{T} L_{L_j} \mathcal{F}
\end{align}
which we have labeled with 'red' since the diagonal elements are redundant. Components with $i\neq j$ are not redundant, and may be CP violating. 

Finally, there are a large number of four-fermion operators.  
Just as in the SMEFT case, we will not list these here, though they may be found in~\cite{Brivio:2016fzo}.

The richness of the HEFT implies that some operators at NLO will not have a direct translation to $d=6$ SMEFT operators. 
A summary of the operators that do have an equivalent can be found in~\cite{Brivio:2016fzo}. 
Since the Higgs basis was originally constructed with the SMEFT in mind, some operators are missing there as well.
In the CP violating case, which is pertinent to us, a detailed explanation is given in Sec. \ref{subsubsec:null-operators} below.

\section{Dictionary for mapping between the Higgs basis, SMEFT and HEFT
\label{subsec:dictionaries}}
In the following subsections, we provide the matching between the CP violating sectors of the Higgs basis, the SMEFT, and the HEFT, using the definitions written above and taking $\Lambda = v$.
The matching between the SMEFT and Higgs basis has been presented before in~\cite{Azatov:2022kbs}, while the matching of the CP violating operators in the HEFT to the Higgs basis has not been presented before to the best of our knowledge.
A subset of the matching of the Higgs basis and the strongly interacting light Higgs (SILH)~\cite{Contino:2013kra} and Hagiwara, Ishihara, Szalapski, Zeppenfeld (HISZ)~\cite{Hagiwara:1993ck} bases are additionally given in~\cite{Falkowski:2001958}.

We should point out that gauge invariance enforces some relations between some of the coefficients in the Higgs basis, regardless of whether the full EFT is SMEFT or HEFT.
This was already used in the quartic gauge couplings in Eq.~\eqref{eq:HiggsBasisQGCs}, and additionally requires $\widetilde{\lambda}_\gamma = \widetilde{\lambda}_z$.

\subsection{SMEFT and the Higgs basis}
The correspondence between the SMEFT in the Warsaw basis and the Higgs basis has been explicitly written in~\cite{Azatov:2022kbs}.
Here we report the pieces relevant for CP violation, following our definitions given above.
First, note that assuming that the Higgs basis comes from the SMEFT in any basis enforces some relations between the Higgs basis coefficients. 
In particular, all operators with two insertions of $H$ must be related to those with only one:
\begin{align}
\begin{split}
\widetilde{c}_{vv}^{(2)} = \widetilde{c}_{vv} , \quad y_f^{(2)} = y_f, \quad d_{Vf} = d_{h V f} , \quad \delta g_{L/R}^{Vf} = \delta g_{L/R}^{hVf} = \delta g_{L/R}^{h^2 Vf}.
\end{split}
\end{align}
We will see this explicitly below.

Starting with the bosonic operators, the translation between the anomalous triple gauge couplings appearing in Eq.\,\eqref{eq:HiggsBasis:tgc} and Eq.\,\eqref{eq:SMEFTbosonic} are given by 
\begin{align}\label{eq:HiggsToSMEFT:TGCs}
\begin{split}
    \widetilde{\kappa}_\gamma = & \ -\frac{g_L}{g_Y}C_{H\widetilde{W}B}\\
    \widetilde{\kappa}_z =& \ \frac{g_Y}{g_L}C_{H \widetilde{W} B}\\
    \widetilde{\lambda}_\gamma = \widetilde{\lambda}_z = & \ -\frac{3}{2}g_L C_{\widetilde{W}}\\
    \widetilde{c}_{3g} =& \ \frac{1}{g_s^3}C_{\widetilde{G}}
\end{split}
\end{align}

The relations between the CP violating Higgs boson-gauge vertices of Eq.\,\eqref{eq:HiggsBasis:hvv} and Higgs boson-fermion vertices of Eq.\,\eqref{eq:HiggsBasis:hff} with the SMEFT operators of Eq.\,\eqref{eq:SMEFTbosonic} and Eq.\,\eqref{eq:SMEFTYukawa} are given by
\begin{align}\label{eq:HiggsToSMEFT:hvv}
\begin{split}
    \widetilde c_{gg} = \widetilde c_{gg}^{(2)} = & \ \frac{4}{g_s^2} C_{H\widetilde{G}} \\
    \widetilde c_{zz} = \widetilde c_{zz}^{(2)} = &\  4\left(\frac{g_L^2 C_{H\widetilde{W}} + g_Y^2 C_{H\widetilde{B}} + g_L g_Y {C_{HW\widetilde{B}}}}{(g_L^2+g_y^2)^2}\right)\\
    \widetilde c_{\gamma \gamma} = \widetilde c_{\gamma\gamma}^{(2)} = & \ 4\left(\frac{C_{H\widetilde{W}}}{g_L^2} -\frac{C_{HW\widetilde{B}}}{g_Y g_L}+\frac{C_{H\widetilde{B}}}{g_Y^2} \right)\\
    \widetilde c_{z \gamma} = \widetilde c_{z \gamma}^{(2)} = &\  \frac{2}{g_Y^2 +g_L^2 }\left(2C_{H\widetilde{W}}-\frac{g_L^2-g_Y^2}{g_L g_Y} C_{HW\widetilde{B}} - 2C_{H\widetilde{B}}\right)\\
    \widetilde c_{ww} = \widetilde c_{ww}^{(2)} = &\  \frac{4}{g_L^2} C_{H\widetilde{W}}\\
    \text{Im}([\delta y_f]_{ij}) = \text{Im}([\delta y^{(2)}_f]_{ij}) = & \ -\frac{v}{\sqrt{2 m_{{f_i}} m_{f_j}}} \text{Im}(C_{fH}^\dagger[ij])
\end{split}
\end{align}
where we have matched only the imaginary parts of the Yukawa matrices to avoid introducing strictly CP-even operators.
We have also included the operators with two Higgs boson insertions from Eq.~\eqref{eq:HiggsBasis:h2} since the matching is identical to the single Higgs boson operators.

Moving on to the dipole operators, the correspondence the operators appearing in Eq.\,\eqref{eq:HiggsBasis:dipoles} and Eq.\,\eqref{eq:SMEFTdipoles} is given simply by 
\begin{align}\label{eq:HiggsToSMEFT:dipoles}
\begin{split}
    d_{Gf}[ij]=d_{hGf}[ij] =& \ -\dfrac{2 \sqrt{2} v}{g_s \sqrt{m_{f_i}m_{f_j}}} C_{fG}[ij]\\
    d_{Af}[ij]=d_{hAf}[ij] =& \ \dfrac{2\sqrt{2} v }{g_L g_Y \sqrt{m_{f_i} m_{f_j}}}\left(g_Y \eta_fC_{fW}[ij]-g_L C_{fB}[ij]\right)\\
    d_{Zf}[ij]=d_{hZf}[ij] =& \ \dfrac{2\sqrt{2} v }{(g_L^2+ g_Y^2) \sqrt{m_{f_i} m_{f_j}}}\left(g_L \eta_fC_{fW}[ij]+g_Y C_{fB}[ij]\right)\\
    d_{We}[ij]=d_{hWe}[ij] =& \ -\dfrac{2\sqrt{2} v }{g_L \sqrt{m_{e_i} m_{e_j}}}\big(U_{\text{PMNS}}^\dagger C_{eW}\big)[ij]\\ 
    d_{Wu}[ij]=d_{hWu}[ij] = &\ -\dfrac{2\sqrt{2} v }{g_L \sqrt{m_{u_i} m_{u_j}}}\big(C_{uW}U_{\text{CKM}}^\dagger\big)[ij]\\
    d_{Wd}[ij]=d_{hWd}[ij] = &\ -\dfrac{2\sqrt{2} v }{g_L \sqrt{m_{d_i} m_{d_j}}}\big(U_{\text{CKM}} C_{dW}\big)[ij]\, .
\end{split}
\end{align}
where we have defined $\eta_u = -1$ and $\eta_{e,d} = +1$ to make the expressions compact.

Finally, the correspondence between the imaginary pieces of the gauge interactions of Eq.\,\eqref{eq:HiggsBasis:vff} and the SMEFT operators of Eq.\,\eqref{eq:SMEFTYukawa} and Eq.\,\eqref{eq:SMEFTgauge} reads
\begin{align}\label{eq:HiggsToSMEFT:gauge}
\begin{split}
    \text{Im}(\delta g_L^{Zu})=\text{Im}(\delta g_{L}^{hZu}) &= \  \dfrac{1}{2}\text{Im}\left[U_\text{CKM} (C_{H q}^{(1)} +C_{H q}^{(3)}) U_\text{CKM}^\dagger\right]\\
    \text{Im}(\delta g_L^{Zd})=\text{Im}(\delta g_{L}^{hZd}) &= \  \dfrac{1}{2}\text{Im}\left[C_{H q}^{(1)} +C_{H q}^{(3)}\right]\\
    \text{Im}(\delta g_L^{Ze})=\text{Im}(\delta g_{L}^{hZe}) &= \  \dfrac{1}{2}\text{Im}\left[C_{H \ell}^{(1)} +C_{H \ell}^{(3)}\right]\\
    \text{Im}(\delta g_R^{Zf})=\text{Im}(\delta g_{R}^{hZf}) &= \  \dfrac{1}{2}\text{Im}\left[C_{H f} \right]\\
    \text{Im}(\delta g_L^{Wq}) = \text{Im}(\delta g_L^{hWq}) &= \  -\text{Im}\left[U_\text{CKM} C_{Hq}^{(3)}\right]\\
    \text{Im}(\delta g_R^{Wq}) = \text{Im}(\delta g_R^{hWq}) &= \  -\dfrac{1}{2}\text{Im}\left[C_{Hud}\right]\\
    \text{Im}(\delta g_L^{W\ell}) = \text{Im}(\delta g_L^{hW\ell}) &= \  \text{Im}\left[U_\text{PMNS}^\dagger C_{H\ell}^{(3)}\right]\\
\end{split}
\end{align}
where we have suppressed flavour indices, though we should emphasize that most of these are Hermitian and can only have CP violating off-diagonal components.

\subsection{HEFT and the Higgs basis}
In this section we present the correspondence between the HEFT and the Higgs basis explicitly for the CPV sector. 
Some comments are in order. 
It is important to note that HEFT at NLO is equipped with more operators than SMEFT at $d=6$, which means that there are some HEFT NLO operators that do not have SMEFT equivalents at $d=6$ but rather at higher dimensions. 
Another detail that must be considered is that in HEFT each operator has an expansion in Higgs fields, which yields different operators 
distinguished by the coefficient accompanying each term of the series expansion in Higgs fields.
We include only up to two Higgs field insertions for each operator since further insertions are not typically relevant for phenomenology. 

Starting with the anomalous triple gauge couplings appearing in Eq.\,\eqref{eq:HiggsBasis:tgc} and Eq.\,\eqref{eq:HEFT-bos}, we have the correspondence
\begin{align}\label{eq:HiggsToHEFT:tgc}
\begin{split}
\widetilde{\kappa}_{\gamma}=& \ -4 \widetilde{c}_{S_{\widetilde{W}}}+8\widetilde{c}_{S_8}+4\dfrac{c_w}{s_w}\widetilde{c}_{S_1}\\   
\widetilde{\kappa}_Z=& \ \dfrac{c_w}{s_w}(-4 \widetilde{c}_{S_{\widetilde{W}}} + 8 \widetilde{c}_{S_8})-4\widetilde{c}_{S_1}\\  
\widetilde{\lambda}_{\gamma} = & \widetilde{\lambda}_z = \  2\pi g_L \widetilde{c}_{S_{\widetilde{W}WW}}\\
    \widetilde c_{3g} =& \  \dfrac{4\pi}{g_s^3}
    \widetilde{c}_{S_{\widetilde{G}GG}}
\end{split}
\end{align}

Moving on to the Higgs boson-gauge operators of Eq.\,\eqref{eq:HiggsBasis:hvv} and Higgs boson-fermions operators of Eq.\,\eqref{eq:HiggsBasis:hff}, the correspondence between the Higgs basis and the HEFT, Eq.\,\eqref{eq:HEFT-bos} and Eq.\,\eqref{eq:HEFTLO}, is the same also for the $H^2$ operators of Eq.\,\eqref{eq:HiggsBasis:h2}, up to a different Wilson coefficient.
In the following we write explicitly only the $ \widetilde{c_j}(a_i)$ coefficients, where $ \widetilde{c_j}(a_i)$ stands for the Wilson coefficient $ \widetilde{c_j}$ as a function of the parameter $a_i$. Notice that the $\widetilde{c}_j^{(2)}(b_i)$ coefficients can be found by replacing $a_i$ with $b_i$ in the definition of the respective $\widetilde{c_j}(a_i)$ coefficient. 

\begin{align}\label{eq:HiggsToHEFT:hvv}
\begin{split}
    \widetilde{c}_{gg}(a_i)= &\ -\dfrac{8 a_{S_{\widetilde{G}}}}{g_S^2} \widetilde{c}_{S_{\widetilde{G}}}\\
    \widetilde{c}_{zz}(a_i) =& \  \dfrac{g_L}{2\pi c_w (g_L^2+g_Y^2)}\bigg(-2 s_w a_{S_2}\widetilde{c}_{S_2}+c_w a_{S_3}\widetilde{c}_{S_3}+2 c_w a_{S_9}\widetilde{c}_{S_9}\bigg)\\
    & \ -\dfrac{8}{g_L^2+g_Y^2}\left( c_w s_w a_{S_1}\widetilde{c}_{S_1}- c_w^2 a_{S_8}\widetilde{c}_{S_8}+s_w^2a_{S_{\widetilde{B}}} \widetilde{c}_{S_{\widetilde{B}}}+\frac{c_w^2}{2} a_{S_{\widetilde{W}}}\widetilde{c}_{S_{\widetilde{W}}}\right)\\
    \widetilde{c}_{\gamma \gamma}(a_i) =& \  -\dfrac{4(g_L^2+g_Y^2)}{g_L^2 g_Y^2}\bigg[2  c_w^2 a_{S_{\widetilde{B}}}\widetilde{c}_{S_{\widetilde{B}}}+s_w^2 a_{S_{\widetilde{W}}} \widetilde{c}_{S_{\widetilde{W}}}\\
    & \quad \qquad \qquad \qquad  -2s_w\left(c_w a_{S_1}\widetilde{c}_{S_1} +s_w a_{S_8}\widetilde{c}_{S_8}\right)\bigg]\\
    \widetilde{c}_{z \gamma}(a_i)=& \  \dfrac{4(c_w^2-s_w^2)}{g_L g_Y}a_{S_1}\widetilde{c}_{S_1}+\dfrac{4 c_w s_w}{g_L g_Y}\left(2 a_{S_8}\widetilde{c}_{S_8}+2a_{S_{\widetilde{B}}} \widetilde{c}_{S_{\widetilde{B}}}-a_{S_{\widetilde{W}}}\widetilde{c}_{S_{\widetilde{W}}}\right)\\
    & \ +\dfrac{1}{2\pi g_Y}a_{S_2}\widetilde{c}_{S_2}+\dfrac{1}{4\pi g_L}\left(a_{S_3}\widetilde{c}_{S_3}+2a_{S_9}\widetilde{c}_{S_9}\right)\\
    \widetilde{c}_{ww}(a_i)=& \   \dfrac{1}{2\pi g_L}a_{S_3} \widetilde{c}_{S_3}-\dfrac{4 }{g_L^2} a_{S_{\widetilde{W}}}\widetilde{c}_{S_{\widetilde{W}}}\\
    \text{Im}([\delta y^{(n)}_f]_{ij}) =& \  \dfrac{v}{\sqrt{2m_{{f_i}} m_{f_j}}} \text{Im}(\mathcal{Y}_f^{(n),\dagger}[ij]) 
\end{split}
\end{align}

The matching of the dipole operators in the case of $d_{Vf}$ coefficients of Eq.\,\eqref{eq:HiggsBasis:dipoles} with the HEFT operators of Eq.\,\eqref{eq:heft_ferm_current_1} and Eq.\,\eqref{eq:heft_ferm_current_2} is given by
\begin{align}\label{eq:HiggsToHEFT:dipoles}
\begin{split}
    d_{Gu}[ij] =& \ \dfrac{16 \pi v}  {\sqrt{m_{u_i} m_{u_j}}}\left(\widetilde{n}_{N_{31}^Q}+\widetilde{n}_{N_{32}^Q}\right)\\
    d_{Gd}[ij] =& \ \dfrac{16 \pi v}  {\sqrt{m_{d_i} m_{d_j}}}\left(\widetilde{n}_{N_{31}^Q}-\widetilde{n}_{N_{32}^Q}\right)\\
    d_{Ae}[ij] =& \ \dfrac{8 \pi s_w v\sqrt{g_L^2+g_Y^2} }{g_Y \sqrt{m_{l_i} m_{l_j}}}\left(2 \widetilde{n}_{N_{12}^\ell}-\widetilde{n}_{N_{13}^\ell}\right)  \\
    d_{Au}[ij] =& \  \dfrac{8 \pi s_w v\sqrt{g_L^2+g_Y^2} }{g_Y \sqrt{m_{u_i} m_{u_j}}}\left(2 \widetilde{n}_{N_{29}^Q}+2 \widetilde{n}_{N_{30}^Q}+\widetilde{n}_{N_{33}^Q}+2 \widetilde{n}_{N_{34}^Q}+\widetilde{n}_{N_{36}^Q}\right) \\
    d_{Ad}[ij] =& \  \dfrac{8 \pi s_w v\sqrt{g_L^2+g_Y^2} }{g_Y \sqrt{m_{u_i} m_{u_j}}}\left(2 \widetilde{n}_{N_{29}^Q}-2 \widetilde{n}_{N_{30}^Q}-\widetilde{n}_{N_{33}^Q}+2 \widetilde{n}_{N_{34}^Q}-\widetilde{n}_{N_{36}^Q}\right)  \\
    d_{Ze}[ij]=& \  -\dfrac{8\pi g_L v}{c_w \sqrt{g_L^2+g_Y^2} \sqrt{m_{l_i} m_{l_j}}}\left(2 \widetilde{n}_{N_{12}^\ell} s_w^2+\widetilde{n}_{N_{13}^\ell} c_w^2\right)\\
    d_{Zu}[ij]=& \  \dfrac{8\pi g_L v}{c_w \sqrt{g_L^2+g_Y^2} \sqrt{m_{u_i} m_{u_j}}}\bigg(-2 \widetilde{n}_{N_{29}^Q} s_w^2-2 \widetilde{n}_{N_{30}^Q} s_w^2+\widetilde{n}_{N_{33}^Q} c_w^2 \\ & \qquad \qquad \qquad \qquad \qquad \qquad +2 \widetilde{n}_{N_{34}^Q} c_w^2+\widetilde{n}_{N_{36}^Q} c_w^2\bigg)\\
    d_{Zd}[ij]=& \  -\dfrac{8\pi g_L v}{c_w \sqrt{g_L^2+g_Y^2} \sqrt{m_{u_i} m_{u_j}}}\bigg(2 \widetilde{n}_{N_{29}^Q} s_w^2-2 \widetilde{n}_{N_{30}^Q} s_w^2+\widetilde{n}_{N_{33}^Q} c_w^2 \\ &\qquad \qquad \qquad \qquad \qquad \qquad -2 \widetilde{n}_{N_{34}^Q} c_w^2+\widetilde{n}_{N_{36}^Q} c_w^2\bigg)\\
    d_{We}[ij]=& \ -\dfrac{8\pi v}{\sqrt{m_{u_i} m_{u_j}}}U_{\text{PMNS}}^\dagger \left(\widetilde{n}_{N_{13}^\ell} - 2 \widetilde{n}_{N_{14}^\ell}\right) \\ 
    d_{Wu}[ij]=& \  \dfrac{8\pi v}{\sqrt{m_{u_i} m_{u_j}}}\left(\widetilde{n}_{N_{33}^Q}+2 \widetilde{n}_{N_{35}^Q}-\widetilde{n}_{N_{36}^Q}\right)U_{\text{CKM}}^\dagger\\
    d_{Wd}[ij] =& \ \dfrac{8\pi v}{\sqrt{m_{d_i} m_{d_j}}}U_{\text{CKM}}\left(\widetilde{n}_{N_{33}^Q}-2 \widetilde{n}_{N_{35}^Q}-\widetilde{n}_{N_{36}^Q}\right).
\end{split}
\end{align}
The relations for the $d_{hVf}$ coefficients can be obtained from the previous equation after replacing $\widetilde{n}_i\to a_i \widetilde{n}_i$ in the respective $d_{Vf}$ equations.

Finally, we have the $Vff$ gauge interactions appearing in Eq.\,\eqref{eq:HiggsBasis:vff} and Eq.\,\eqref{eq:HEFT-2f}
\begin{align}\label{eq:HiggsToHEFT:gauge}
\begin{split}
    \text{Im}(\delta g_L^{Zu}) &= \ -\dfrac{1}{2}\text{Im}\left[U_\text{CKM} (n_{N_{1}^Q}+2n_{N_5^Q}+n_{N_7^Q}) U_\text{CKM}^\dagger\right]\\
    \text{Im}(\delta g_L^{Zd}) &= \  \dfrac{1}{2}\text{Im}\left[(n_{N_{1}^Q}-2n_{N_5^Q}+n_{N_7^Q}) \right]\\
    \text{Im}(\delta g_L^{Ze}) &= \  \dfrac{1}{2}\text{Im}\left[(n_{N_{\text{red},1}^\ell}-2n_{N_{\text{red},2}^\ell}+n_{N_{\text{red},3}^\ell})\right]\\
    \text{Im}(\delta g_R^{Zu}) &= \  -\dfrac{1}{2}\text{Im}\left[(n_{N_2^Q}+2n_{N_6^Q}+n_{N_8^Q})\right]\\
    \text{Im}(\delta g_R^{Zd}) &= \  \dfrac{1}{2}\text{Im}\left[(n_{N_2^Q}-2n_{N_6^Q}+n_{N_8^Q})\right]\\
    \text{Im}(\delta g_R^{Ze}) &= \  -\text{Im}\left[n_{N_2^\ell}\right]\\
    \text{Im}(\delta g_L^{Wq}) &= \  -\text{Im}\left[U_\text{CKM} (n_{N_1^Q} -n_{N_7^Q}+ 2 i n_{N_3^Q})\right]\\
    \text{Im}(\delta g_R^{Wq}) &= \  -\text{Im}\left[(n_{N_2^Q} -n_{N_8^Q}+ 2 i n_{N_4^Q})\right]\\
    \text{Im}(\delta g_L^{W\ell}) &= \ -\text{Im}\left[U_\text{PMNS}^\dagger (n_{N_{\text{red},1}^\ell} - n_{N_{\text{red},3}^\ell}-2 i n_{N_1^\ell})  \right]\\
\end{split}
\end{align}
where once again the $\delta g_{L/R}^{hVf}$ coefficients may be found with the replacement $\widetilde{n}_i\to a_i \widetilde{n}_i$.

\subsection{Operators unique to the HEFT}
\label{subsubsec:null-operators}
Due to the fact that the Higgs boson is a singlet in the HEFT Lagrangian, there are some operators which are allowed in the HEFT but not in the SMEFT. 
These operators would require an extension of the Higgs basis Lagrangian to include in the matching.
For example, operators that give rise to triple gauge coupling with a Higgs boson will not be present in SMEFT, since the $SU(2)_L$ symmetry will be broken.
These would require the additional Higgs basis Lagrangian written in Eq.~\eqref{eq:HiggsBasis:htgc}.
If we wish to instead stick to the original definition of the Higgs basis Lagrangian, these operators must have their coefficients set to zero in the matching.
The other operators that are present in Eq.\,\eqref{eq:HEFT-bos} and Eq.\,\eqref{eq:HEFT-2f} but that are not present in the dictionary also do not have a SMEFT comparison.
A full list of the CP violating operators unique to the HEFT is as follows:
\begin{center}
\begin{tabular}{ c|c } 
 Bosonic operators &$a_{S_{\widetilde{X}XX}}, b_{S_{\widetilde{X}XX}}$ for $X=W,G$, $ \quad\widetilde{c}_{S_i}$, with $i=4,...,7,15$ \\ 
 2-quark operators & $\widetilde{n}_{N_{i}^Q}\,,\,\text{with } i=9,.., 28$ \\ 
 2-lepton operators & $\widetilde{n}_{N_{i}^\ell}\,,\,\text{with } i=3,..,11$ \\ 

\end{tabular}
\end{center}

\section{Conclusions\label{sec:conclusions}}

The search for CP violating interactions in the Higgs sector remains a key objective in the LHC program and a crucial avenue for testing the SM and probing physics beyond it. Additional sources of CP violation are in fact a necessary condition for realizing the Universe baryon asymmetry. The ATLAS and CMS experiments at the LHC have performed several analysis for the direct search of CP violating Higgs boson couplings. With the forthcoming HL-LHC dataset the experimental sensitivity to CP violating effects in Higgs production and decay channels will substantially improve. These measurements are expected to provide the most stringent direct limits on CP violation in the Higgs sector for the foreseeable future, setting a solid foundation for future collider explorations and serving as a benchmark for the interpretation of new physics effects. However, a precise and unambiguous interpretation of experimental results requires a consistent treatment of the various theoretical frameworks and parametrisations used to describe CPV effects.

In this paper, we have reviewed the most commonly employed parametrisations of Higgs interactions relevant to CP violation, namely the Higgs basis, the $\kappa$ and angles framework, CP fractions and effective field theories such as the SMEFT and HEFT and constructed a dictionary that translates between these parametrisations.
This includes explicit relations between Wilson coefficients in SMEFT and operator coefficients in the Higgs basis and HEFT, highlighting the regimes where the mappings are valid or break down. Special attention has been devoted to identifying operator redundancies, the so-called "null operators" in HEFT that do not contribute to physical observables and the underlying flavor structure.

For parametrisations often used in experimental analyses, we have shown how $\kappa$'s, angles, and CP fractions are equivalent to one another and to measuring ratios of EFT coefficients. 
These parameterizations on their own may always be used independent of any questions of EFT operator dimensionality or validity.
To interpret them in a typical EFT fit, however, the experimental precision must be sufficient to be driven by the linear dimension-6 interference term only, with negligible difference when including quadratic or dimension-8 contributions. 
If not, then it cannot be included in the EFT fit, though the measurement and analysis remains perfectly valid in its original parameterization.
In reporting results in these parametrisations, it is helpful to report typical energy ranges used in an analysis (i.e. $p_T$ or invariant mass ranges that drive the sensitivity), which help determine when an EFT fit is valid, as well as any assumptions that are made, if present.
If results are given in an EFT interpretation, it is especially important to show the impact of including and neglecting quadratic contributions to ensure validity.

To put our results in perspective, we also briefly summarized
some of the recent literature relevant for CP violation in the Higgs sector,
emphasizing how different parametrisations impact the interpretation of the results.

Our results aim to provide a unifying language to bridge different theoretical approaches and experimental strategies, promoting a more transparent and accurate comparison of CP violating studies in the Higgs sector. This can be particularly valuable in global fits, reinterpretation of results, and the design of future analyses both at the LHC and future colliders.

\section*{Acknowledgements}
\paragraph{Funding information}
M.F. is supported by the U.S. National Science Foundation grant PHY-2210533.
The work of M.F.Z. is supported by the Spanish MIU through
the National Program FPU (grant number FPU22/03625) and by the grant PID2022-137127NB-I00 funded by MCIN/AEI/ 10.13039/501100011033.
A.G. gratefully acknowledges support from the U.S. NSF under grant PHY-2310072.
P.P.G. and M.F.Z. are supported by the Spanish Research Agency (Agencia Estatal de Investigación) through the grant IFT Centro de Excelencia Severo Ochoa~No~CEX2020-001007-S. 
P.P.G. is supported by the Ramón y Cajal grant~RYC2022-038517-I funded by MCIN/AEI/10.13039/501100011033 and by FSE+. 

\newpage
{\footnotesize
\bibliography{biblio}}
\end{document}